\documentclass[aps,pra,twocolumn,floatfix]{revtex4-1}

\usepackage{graphicx}
\usepackage[english]{babel}
\usepackage{amsmath}
\usepackage{amssymb}

\allowdisplaybreaks
\begin{document}
\title{Shear Viscosity of Uniform Fermi Gases with Population Imbalance}

\author{Weimin Cai$^{1}$, Yan He$^{2}$, Hao Guo$^{1}$, Chih-Chun Chien$^{3}$}
\affiliation{$^1$Department of Physics, Southeast University, Nanjing 211189, China}
\affiliation{$^2$College of Physical Science and Technology, Sichuan University, Chengdu, Sichuan 610064, China}
\affiliation{$^3$School of Natural Sciences, University of California, Merced, CA 95343, USA}
\email{cchien5@ucmerced.edu}

\begin{abstract}
The shear viscosity plays an important role in studies of transport phenomena in ultracold Fermi gases and serves as a diagnostic of various microscopic theories.
Due to the complicated phase structures of population-imbalanced Fermi gases, past works mainly focus on unpolarized Fermi gases. Here we investigate the shear viscosity of homogeneous, population-imbalanced Fermi gases with tunable attractive interactions at finite temperatures by using a pairing fluctuation theory for thermodynamical quantities and a gauge-invariant linear response theory for transport coefficients. In the unitary and BEC regimes, the shear viscosity increases with the polarization because the excess majority fermions cause gapless excitations acting like a normal fluid. In the weak BEC regime the excess fermions also suppress the noncondensed pairs at low polarization, and we found a minimum in the ratio of shear viscosity and relaxation time.  To help constrain the relaxation time from linear response theory, we derive an exact relation connecting some thermodynamic quantities and transport coefficients at the mean-field level for unitary Fermi superfluids with population imbalance. An approximate relation beyond mean-field theory is proposed and only exhibits mild deviations from numerical results.
\end{abstract}

\maketitle


\section{Introduction}
Ultracold Fermi gases provide versatile quantum simulators for complex many-particle systems, and  their transport properties have attracted
broad research interest \cite{Son1,KinastPRL05,BruunPRA07,SchaferPRA07,ourlongpaper,ThomasJLTP08,Nascimbene10,EnssPRA12,ThomasPRL14,ThomasScience11,PethickPRL11,ZwierleinNature11,ProchePRA13,SchaeferPRA14,YanPRB14,SchaferPRA15,ThomasPRL15,SchaeferPRL16}. Among those transport phenomena, the shear viscosity relating the momentum transfer transverse to a shear force is often thought of as an important diagnostic of various microscopic or phenomenological theories for many-body systems. In particular, the ratio of the shear viscosity to entropy density of unitary Fermi gases, where fermions are about to form two-body bound states, has been shown to be close to the quantum lower bound~\cite{TDSonPRL05,Turlapov08,Cao11}. Previous theoretical works, however, mainly focus on the shear viscosity of two-component Fermi gases with equal populations~\cite{HaoPRL11,EnssPRA12,HaoNJP11,SchaferPRA15}.

Since the populations of different components of ultracold Fermi gases can be adjusted~\cite{ZSSK06,ZSSK206,ZSSK07,MITPRL08,Rice06,RicePRL06}, 
population-imbalanced Fermi gases have been an intensely studied subject in cold-atoms. The corresponding studies of population-imbalanced Fermi gases are more difficult since the low temperature phase structure is not a homogeneous mixture but rather a phase separation of paired and unpaired fermions~\cite{Rice06,ZSSK07}. 
For example, the ``intermediate-temperature superfluid'' describes where homogeneous polarized superfluids appear~\cite{Chien06} when a population-imbalanced ultracold Fermi gas undergoes the BCS-Bose Einstein Condensation (BEC) crossover as the attractive interaction increases. Here we focus on 3D systems and mention that other exotic phases and structures may emerge in 1D population-imbalanced Fermi gases~\cite{Liao10}.

In the unitary and BEC regimes, preformed pairs due to strongly attractive interactions lead to noncondensed pairs which do not contribute to superfluidity. Theories incorporating noncondensed pairs are often called pairing-fluctuation theories~\cite{Ourreview,NSR,ZwergerPRA07}. 
At finite temperatures, the noncondensed pairs can lead to an energy gap in the single-particle dispersion even in the absence of superfluidity, which is usually called the pseudogap~\cite{OurNSR10}.
Thus, the pairing energy gap $\Delta$ should be distinguished from the order parameter $\Delta_{\textrm{sc}}$ describing the condensed, coherent Cooper pairs. As a consequence, the pairing onset temperature $T^*$ is higher than the superfluid
transition temperature $T_c$ in the strongly attractive regime.
Among various approaches, one particular pairing-fluctuation theory consistent with the Leggett-BCS theory has been successfully generalized to describe polarized Fermi gases in the BCS-BEC crossover~\cite{Chien06,ChienPRL}, and we will implement this particular theory here. The equations of state of population-imbalanced Fermi gases can be obtained and thermodynamic quantities such as the chemical potential and pressure can be determined.

A theoretical study of the shear viscosity of ultracold Fermi gases
with population imbalance will be presented here, based on a gauge-invariant linear response theory called the consistent fluctuation of oder parameter (CFOP) theory~\cite{OurJLTP13}.
The theoretical framework and key results are summarized in the Appendix.
Importantly,
a consistent description of the shear viscosity beyond mean-field should include both fermionic and bosonic (from noncondensed pairs) contributions~\cite{OurPRA17}. Here, the former is obtained by the
CFOP theory \cite{OurPRD12,OurJLTP13} while the latter is approximated by an effective bosonic theory~\cite{OurPRA17}. Since the system is phase separated at low temperatures in the BCS regime, we focus our studies on the unitary and BEC regimes at finite temperatures.

To help constrain the elusive relaxation time which appears in the transport coefficients obtained from linear response theory, we generalize a relation~\cite{OurPRA17} connecting thermodynamic quantities and transport coefficients previously derived for unpolarized unitary Fermi superfluids. An exact relation for homogeneous, population-imbalanced unitary Fermi superfluids at the mean-field level will be presented. However, pairing fluctuations make a full derivation of the relation beyond mean-field quite complicated, and we instead present an approximate relation including the contributions from noncondensed pairs.

The paper is organized as follows. In Sec.~\ref{sec:EOS} we briefly review the mean-field and pairing-fluctuation theories along with the corresponding gauge-invariant linear response theory for population-imbalanced Fermi gases in the BCS-BEC crossover. Sec.~\ref{sec:viscosity} presents the shear viscosity from mean-field and pairing-fluctuation calculations. For the latter, the fermionic and bosonic contributions are evaluated and discussed. Numerical results and analyses are presented subsequently.
To help determine the relaxation time, in Sec.~\ref{sec:relation} we present a relation connecting the pressure, chemical potential, shear viscosity, superfluid density, and anomalous shear viscosity of polarized unitary Fermi superfluids at the mean-field level. Then, an approximate relation is proposed and numerical results show the approximation works reasonably. Sec.~\ref{sec:conclusion} concludes our study.
The theoretical details and derivations are summarized in the Appendix.

\section{Mean-field and beyond mean-field theories of polarized Fermi gases}\label{sec:EOS}
At the mean-field level, the equations of state of a two-component (labeled by $\sigma=\uparrow,\downarrow$) population-imbalanced Fermi gas include two number equations and a gap equation~\cite{ourlongpaper,Pao,StrinatiPRL06,HuiHuEPL06}. Assuming the two components have the same fermion mass $m$ and densities $n_{\uparrow,\downarrow}$, the equations are given by
\begin{eqnarray}\label{eos}
n&\equiv&n_\uparrow+n_\downarrow=\sum_{\mathbf{k}}\Big[1-\frac{\xi_{\mathbf{k}}}{E_{\mathbf{k}}}\big(1-2\bar{f}(E_{\mathbf{k}})\big)\Big],\nonumber\\
\delta n&\equiv&n_\uparrow-n_\downarrow=\sum_{\mathbf{k}}\big(f(E_{\mathbf{k}\uparrow})-f(E_{\mathbf{k}\downarrow})\big),\nonumber\\
\frac{1}{g}&=&\sum_{\mathbf{k}}\frac{1}{2\epsilon_{\mathbf{k}}}-\frac{m}{4\pi a}=\sum_{\mathbf{k}}\frac{1-2\bar{f}(E_{\mathbf{k}})}{2E_{\mathbf{k}}},
\end{eqnarray}
where $\mu=\frac{\mu_\uparrow+\mu_\downarrow}{2}$, $h=\frac{\mu_\uparrow-\mu_\downarrow}{2}$, $\epsilon_\mathbf{k}=k^2/2m$, $\xi_\mathbf{k}=\epsilon_\mathbf{k}-\mu$, $E_\mathbf{k}=\sqrt{\xi^2_\mathbf{k}+\Delta^2}$ , $E_{\mathbf{k}\uparrow,\downarrow}=E_\mathbf{k}\mp h$,  $f(x)=1/(1+e^{x/T})$ is the Fermi distribution function, and $\bar{f}(x)=\big(f(x+h)+f(x-h)\big)/2$. There is no distinction between the order parameter and single-particle energy gap at the mean-field level, so we use $\Delta$ to denote the gap. Here we take the convention $\hbar=1$, $k_B=1$, $K=(i\omega_n,\mathbf{k})$, $Q=(i\Omega_l,\mathbf{q})$ and $\sum_K=T\sum_{\omega_n}\sum_\mathbf{k}$ where $\omega_n$ ($\Omega_l$) is the fermionic (bosonic) Matsubara frequency. Moreover, $\mu_\sigma$ is the chemical potential for each component (spin), $g$ is the attractive coupling constant
modeling the contact interaction between atoms,
and $a$ is the two-body $s$-wave scattering length.
The unitary limit is determined by $1/(k_Fa)=0$, where $k_F$ is the Fermi momentum of a noninteracting Fermi gas with the same density.

As the attractive interaction increases, an unpolarized Fermi gas undergoes the BCS-BEC crossover~\cite{Leggett}, where the ground state changes from a collection of Cooper pairs to a condensate of composite bosons. The BCS (BEC) regime corresponds to $1/(k_F a) <0$ ($1/(k_F a) >0$). In contrast, the ground state of a polarized Fermi gas can exhibit structural transitions and a phase separation of paired and unpaired fermions can emerge in the BCS and unitary regimes~\cite{ChienPRL,ourlongpaper,Chienthesis}.

On the other hand, preformed pairs not contributing to the superfluid start to form at finite temperatures as the attractive interaction gets stronger, and we consider the more realistic situation which includes pairing fluctuation effects \cite{ourlongpaper,OurAnnPhys,Ourreview,OurNSR10,NSR}.
Here we follow a particular scheme~\cite{Ourreview,Chien06} consistent with the BCS-Leggett ground state in the unpolarized limit. In this theory, the full Green's function is $G_\sigma(K)=[G_{0\sigma}(K)-\Sigma_\sigma(K)]^{-1}$. Here $G_{0\sigma}(K)=(i\omega_n-\xi_{\mathbf{k}\sigma})^{-1}$ is the bare Green's function with $\xi_{\mathbf{k}\sigma}=\epsilon_\mathbf{k}-\mu_\sigma$ and the fermion self-energy is $\Sigma_\sigma(K)=\sum_{Q}t(Q)G_{0\bar{\sigma}}(Q-K)$ with $\bar{\sigma}$ being the opposite of $\sigma$. To construct the $t$-matrix, we consider
a spin-symmetrized pair susceptibility, or one rung of the ladder diagrams, consisting of one bare and one full Green's functions with the expression  $X(Q)=\frac{1}{2}\sum_K[G_{0\uparrow}(Q-K)G_\downarrow(K)+G_{0\downarrow}(Q-K)G_\uparrow(K)]$.
The $t$-matrix $t(Q)$ is separated into the condensed (sc, $Q=0$) and noncondensed (pg, $Q\neq 0$) pair contributions as $t(Q)\approx t_\textrm{sc} + t_\textrm{pg}$ with $t_{\textrm{sc}}(Q)=-(\Delta_{\textrm{sc}}^2/T)\delta(Q)$ and $t_\textrm{pg}(Q)=[g^{-1}+X(Q)]^{-1}$. The gap function also has two contributions $\Delta^2(T)=\Delta^2_\textrm{sc}(T)+\Delta^2_\textrm{pg}(T)$ from the condensed and
noncondensed pairs. Here $\Delta_\textrm{sc}$ is the order parameter and $\Delta_\textrm{pg}$ is the pseudogap which is approximated
by $\label{pg}\Delta_\textrm{pg}^2\approx-\sum_{Q\neq 0}t_{\textrm{pg}}(Q)$. The pairing onset temperature is determined by the temperature at which the total gap $\Delta$ vanishes, while the superfluid transition temperature is determined by where the order parameter $\Delta_\textrm{sc}$ vanishes.

The equations of state with pairing fluctuations can be derived from $n=\sum_{K,\sigma}G_\sigma(K)$, $\delta n=\sum_K\big(G_\uparrow(K)-G_\downarrow(K)\big)$ and $\frac{1}{g}=\frac{1}{2}\sum_{K,\sigma}G_\sigma(K)G_{0\bar{\sigma}}(-K)$. Their explicit expressions are formally the same as Eqs.~(\ref{eos}), but one has to distinguish the order parameter from the total gap. Similar to the mean-field result, at low temperatures the polarized Fermi gas is unstable against phase separation in the BCS and unitary regimes \cite{Chien06,ZSSK07}.
Here we focus on the homogeneous phases and leave the definition and investigation of shear viscosity in the phase separated structures for future studies.

\begin{figure}[t]
\centering
\includegraphics[width=3.4in, clip]{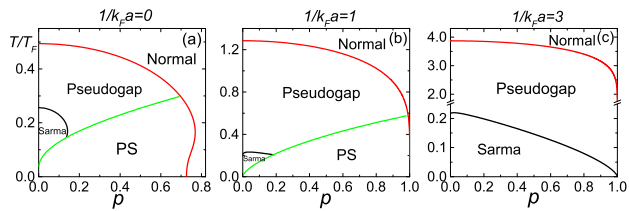}
 \caption{$T$-$p$ phase diagrams of population-imbalanced Fermi gases in a box potential (a) in the unitary limit ($1/k_F a=0$) and (b), (c) in the BEC regime ($1/k_Fa = 1, 3$). 'Sarma' indicates the uniform superfluid
phase, 'PS' corresponds to the phase separation, 'Pseudogap' denotes the homogeneous paired normal
phase, and 'Normal' denotes the unpaired Fermi gas phase. Here $n=k^3_F/3\pi^2$, $E_F=k_BT_F=\hbar^2k^2_F/2m$.  }
 \label{fig.0}
\end{figure}

The phase diagrams of polarized Fermi gases have been shown in Refs.~\cite{Chien06,ChienPRL} for gases in box potentials and harmonic traps. Fig.~\ref{fig.0} shows the phase diagrams of polarized Fermi gases in a box potential in the unitary ((a) for $1/k_Fa=0$) and BEC ((b) for $1/k_Fa=1$ and (c) for $1/k_Fa=3$) regimes. The polarization is defined by $p=\delta n/n$. The homogeneous superfluid or pseudogap phase in the BCS and unitary regimes is unstable at low temperatures against phase separation. The phase separation (PS) structure is a coexistence of an unpolarized superfluid or pseudogap normal gas made of fermion pairs and a normal gas made of excess fermions~\cite{Chienthesis}.

To locate where phase separation emerges at low temperatures, we adopt a simplified approach: the unpaired normal phase has a fraction $x$ of the total particles while the paired phase has a fraction $1-x$, and the two phases are separated by an interface with positive energy. Since the system is in equilibrium, $T$, $\mu_\sigma$ and $P$ should be continuous across the interface~\cite{Caldas03_04}.
The phase boundary between a stable polarized superfluid phase (called the Sarma phase~\cite{Sarma}) or pseudogap phase and the phase separation is given by the condition $x=0$.
On the deep BEC side, the pairing gap is large and the polarized superfluid phase is robust, so there is no PS even at low temperatures.
Since we focus on the shear viscosity in homogeneous phases, we will address the regimes besides PS shown in Fig.\ref{fig.0}.

\section{Shear Viscosity in Polarized Fermi Gases}\label{sec:viscosity}
The shear viscosity can be evaluated by linear response theory, or the Kubo formalism~\cite{Kadanoff61,OurPRA17},
\begin{eqnarray}\label{eta0}
\eta=-m^2\lim_{\omega\rightarrow0}\lim_{q\rightarrow0}\frac{\omega}{q^2}\textrm{Im}K_{\textrm{T}}(Q),
\end{eqnarray}
where the transverse current-current correlation function is defined by $K_\textrm{T}=(\sum_{i=x}^z K^{ii}_{\textrm{JJ}}-K_{\textrm{L}})/2$ with the longitudinal part given by
$K_\textrm{L}=\hat{\mathbf{q}}\cdot\tensor{K}_{\textrm{JJ}}\cdot\hat{\mathbf{q}}$. The frequency is obtained by a complex continuation of the bosonic Matsubara frequency $i\Omega_l\rightarrow\omega+i0^+$, so $Q$ becomes $(\omega,\mathbf{q})$ and $\hat{\mathbf{q}}=\mathbf{q}/|\mathbf{q}|$. The current-current response function can be obtained from the gauge-invariant CFOP theory summarized in Appendix~\ref{appeta}.

Similar to the decomposition of the total energy gap, the shear viscosity of strongly interacting Fermi gases receives contributions from the condensed and noncondensed fermion pairs and fermionic quasiparticles. Thus, $\eta=\eta_\textrm{f}+\eta_\textrm{b}$. Here the subscripts ``f'' and ``b'' represent the condensed-pair (plus fermionic-quasiparticle) and noncondensed-pair contributions respectively. The former can still be obtained from the CFOP linear response theory via Eq.~(\ref{eta0}),
where
the response function $K^{ij}$ can be obtained from Eq.~(\ref{Kmn}).
The bosonic contribution will be discussed later. To ensure the consistency between the thermodynamics and response functions, it is important to find a gauge invariant vertex satisfying the Ward identity (\ref{WI0}), which is also addressed in Appendix~\ref{appb1}.

When the attractive interaction becomes stronger, finding an explicit expression for a gauge invariant vertex is difficult since the vertex must be modified in the same way as the self-energy in the Green's function~\cite{Nambu60}.  After incorporating the relaxation time from linear response theory, the fermionic part of the shear viscosity, from the condensed pairs and fermionic quasiparticles, is
\begin{eqnarray}\label{SVf}
\eta_\textrm{f}&=&\frac{1}{30\pi^2m^2}\int_0^{\infty}dkk^6\Big(1-\frac{\Delta^2_{\textrm{pg}}}{E^2_{\mathbf{k}}}\Big)\nonumber\\
&\times&\frac{\xi^2_{\mathbf{k}}}{E^2_{\mathbf{k}}}\Big[-\frac{\partial f(E_{\mathbf{k}\uparrow})}{\partial E_{\mathbf{k}\uparrow}}-\frac{\partial f(E_{\mathbf{k}\downarrow})}{\partial E_{\mathbf{k}\downarrow}}\Big]\tau,
\end{eqnarray}
The details of its derivation can be found in Appendix.~\ref{appb1}. We emphasize that in Eq.~(\ref{SVf}) there are also contributions from bosonic excitations via the terms involving  $\Delta^2_\textrm{pg}$, which reflects a reduction of the fermionic normal fluid due to strong pairing effect.

The bosonic contribution $\eta_\textrm{b}$ comes from the noncondensed pairs which are approximated by a noninteracting Bose gas with renormalized mass and chemical potential in our theory.
For numerical calculations, the $t$-matrix is approximated by~\cite{LOFFlong} 
$t_\textrm{pg}(\Omega,\mathbf{q})\approx\frac{1}{a_0(\omega-\Omega_\mathbf{q})}$. Here $a_0=\frac{\partial \chi(Q)}{\partial \Omega}|_{Q=0}$ and $\Omega_\mathbf{q}=\frac{q^2}{2M^*}-\mu_\textrm{pair}$ with $M^*=12\frac{\partial^2 \chi(Q)}{\partial q^2}|_{Q=0}$ being the effective pair mass and $\mu_\textrm{pair}$ the pair chemical potential. $\mu_\textrm{pair}$ is negative since it accounts for the binding energy of fermion pairs. The pseudogap is then approximated by $\Delta^2_\textrm{pg}\approx a^{-1}_0\sum_\mathbf{q}b(\Omega_\mathbf{q})$, where $b(x)=1/(\exp(x/T)-1)$ is the Bose distribution function.
Then, $\eta_\textrm{b}$ is evaluated by approximating the noncondensed pairs as noninteracting bosons with energy dispersion $\Omega_\mathbf{q}$. The bosonic contribution to the shear viscosity is given by
\begin{align}\label{SV2}
\eta_\textrm{b}=-\frac{1}{30\pi^2M^{\ast2}}\int_0^\infty dkk^6\frac{\partial b(\Omega_\mathbf{k})}{\partial \Omega_\mathbf{k}}\tau.
\end{align}
An outline of the derivation can be found in Appendix~\ref{appb1}, and the expression for unpolarized Fermi gases is given in Ref.~\cite{OurPRA17}. Here we have assumed that the relaxation time $\tau$ of composite bosons is the same as
that of the fermions since the noncondensed pairs are in local equilibrium
with the fermions.

\begin{figure}[th]
\centering
\includegraphics[width=0.8\columnwidth, clip]{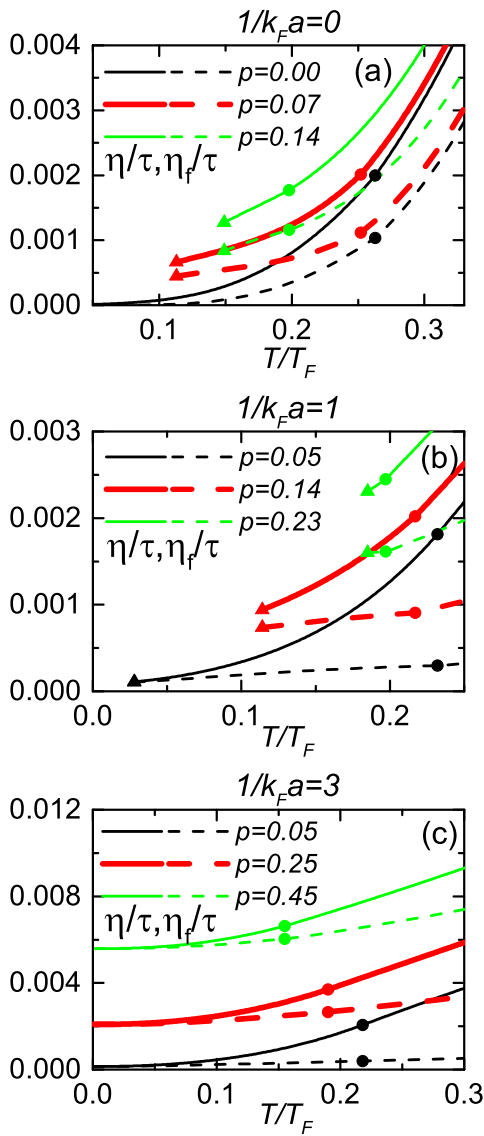}
 \caption{Shear viscosity (in units of $E_Fk^3_F$) as a function of temperature at unitarity (panel (a)) and the BEC side (panel (b)) for different $p$.  The triangles denote where the homogeneous phase becomes unstable and phase separation occurs. The circle denotes where $T_c$ is. The solid lines correspond to $\eta/\tau$, and the dashed lines correspond to $\eta_\textrm{f}/\tau$.}
 \label{fig.1}
\end{figure}

The relaxation time $\tau$ may be obtained by the Boltzmann
equation approach at high temperatures~\cite{Conti,Dorfle80,BruunPRA05} 
 or via the approximate relation $\tau\approx-2\textrm{Im}\Sigma(K)$ \cite{BruunPRA07}, where $\Sigma$ is the self-energy of fermionic quasi-particles.
However, a fully consistent formalism is still lacking for the former at low temperatures when fermion pairs are present.
For the latter, the analytical structure of the self-energy is complicated at low temperatures,
and a first-principle numerical treatment below $T_c$ remains a challenge. Here we first focus on the ratio  $\eta/\tau$ and will later present a relation which may help determine the elusive $\tau$ for polarized unitary Fermi gases.

In our numerical calculations, we fix the total particle density $n$.
In Fig.~\ref{fig.1}, we show $\eta/\tau$ and $\eta_\textrm{f}/\tau$ as a function of temperature for polarized Fermi gases in the unitary and BEC regimes
with selected polarization $p$. In the unitary limit shown in panel (a),
the polarization is restricted to $0\le p\lesssim 0.14$ where superfluids exist (see Fig.~\ref{fig.0} (a)).
Both $\eta/\tau$ and $\eta_\textrm{f}/\tau$ increase as the temperature increases since the number of condensed pairs decreases due to thermal
excitations. The behavior of $\eta_\textrm{b}/\tau$ can be obtained by $(\eta-\eta_\textrm{f})/\tau$, and it also increases with temperature since the number of noncondensed pairs treated as a normal bosonic gas
increases with $T$ below the pairing onset temperature $T^*$. In the BEC regime, illustrated in panel (b), the polarization is restricted to $p\lesssim0.21$ where superfluids survive at low temperatures.
Only when $1/k_Fa\gtrsim2.3$ in the deep BEC regime, the system becomes fully stable against phase separation.
Panel (c) shows the case of $1/k_Fa=3.0$ and the superfluid phase is stable at low temperatures.
Although the basic trend is similar to panel (b), $\eta_\textrm{b}/\tau$ contribute more significantly as temperature increases. This is because the noncondensed pairs in the deep BEC regime behave like thermal bosons, whose fraction increases with temperature.

\begin{figure}[th]
\centering
\includegraphics[width=0.8\columnwidth, clip]{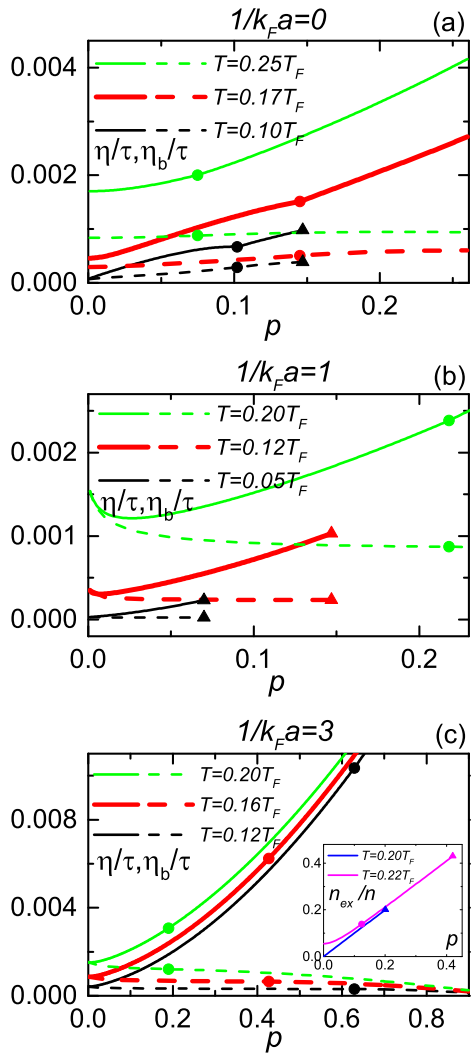}
 \caption{Shear viscosity as a function of $p$ in the unitary limit (panel (a)), BEC side (panel (b)) and deep BEC side (panel (c)) at different temperatures. The convention is the same as that of Fig.~\ref{fig.1} except the dashed line denotes $\eta_\textrm{b}/\tau$. The inset of panel (c) shows the number density of gapless excitations as a function of $p$. The red (black) line indicates the situation with $T=0.22T_F$ and $1/(k_Fa)=0$ at unitarity ($T=0.20T_F$ and $1/(k_Fa)=1$ in the BEC regime).
 }
 \label{fig.2}
\end{figure}

To better understand the dependence of the shear viscosity on the polarization,
we show $\eta/\tau$ and $\eta_\textrm{b}/\tau$ (instead of $\eta_\textrm{f}/\tau$) as a function of $p$
from low to high temperatures in the unitary and BEC regimes in Figure~\ref{fig.2}.
The increase of $\eta/\tau$ with $p$ is quite noticeable because the population-imbalanced superfluid is a homogeneous mixture of the condensed pairs and excess majority fermions, and the latter cause gapless excitations acting like a normal fluid which leads to finite shear viscosity (elaborated below). Therefore, as the polarization increases, the relative population of condensed pairs decreases, and the shear
viscosity increases.

At unitarity,
the noncondensed pair contribution $\eta_\textrm{b}/\tau$ at low temperatures shows a relatively upward trend as $p$ increases, but it saturates at higher temperatures. This trend is opposite to that in the BEC regimes shown in panels (b) and (c).
This is because the pairing gap at unitarity is smaller compared to the gap in the deep BEC regime, and the properties of condensed and noncondensed pairs depend more sensitively on temperature and polarization at unitarity. While the effective mass of pairs, $M^*$, approaches $2m$ in the BEC regime because the fermions are tightly bound, we found $M^*$ increases with $p$ at unitarity.
In Eq.~(\ref{SV2}), $M^*$ appears both in the denominator and the bosonic dispersion $\Omega_\mathbf{q}$ and the combined effect causes the upward trend of $\eta_\textrm{b}/\tau$ as $p$ increases at low temperatures in the unitary limit. As the system enters the BEC regime, $M^*$ no longer increases with $p$ and $\eta_\textrm{b}/\tau$ decreases with $p$ due to a decreasing fraction of paired fermions.

Fig.~\ref{fig.2} (c) shows the result in the deep BEC regime. Since the strongly attractive interactions allow the superfluid and pseudogap phases to be highly polarized and accommodate excess fermions, the bosonic contribution $\eta_\textrm{b}$ becomes less dominant as $p$ increases. Moreover, thermal excitations are also less prominent because both the noncondensed pairs and excess fermions have smooth thermal distributions. The shear viscosity of polarized Fermi gases comes mainly from the gapless excitations caused by the excess fermions. This can be understood by the 
energy dispersion $E_{\mathbf{k}\uparrow,\downarrow}=E_\mathbf{k}\mp h$ of the fermionic excitations. If $n_\uparrow >n_\downarrow$, $h>0$ and $E_{\mathbf{k}\uparrow}<0$ if $k\in [k_1, k_2]$ and $\mu^2+\Delta^2\ge h^2$ or if $k\in [0, k_2]$ and $\mu^2+\Delta^2 < h^2$ , where $k_{1,2}=\sqrt{2m\mu\mp 2m\sqrt{h^2-\Delta^2}}$. In both cases, the excitations are gapless because of the excess fermions.
The contribution from the gapless excitations can be estimated by $n_\textrm{ex}=\sum_{\mathbf{k},\sigma}f(E_{\mathbf{k},\sigma})$, which only takes significant values if the dispersion is gapless and counts the number of fermionic excitations. In both unitary and BEC regimes, $n_\textrm{ex}$ increases with $p$  as shown in the inset of Fig~\ref{fig.2}(c). The excitations behave like a normal fluid and dominate the contribution to the shear viscosity at higher $p$.

Fig.~\ref{fig.2} (b) shows the results
in the shallow BEC regime with $1/k_Fa=1.0$. Interestingly, here the noncondensed pairs contribute more significantly to the shear viscosity at low $p$. One can see that
indeed $\eta\simeq\eta_b$ at low $p$ because condensed pairs form a superfluid and do not contribute to the shear viscosity, so the contribution is mostly from the noncondensed pairs behaving like a normal fluid.
At higher temperatures, interestingly, $\eta/\tau$ is not monotonic as $p$ increases and a minimum emerges. This is because the number of fermion pairs, including both condensed and noncondensed pairs, decreases as $p$ or $T$ increases as more excess fermions or fermionic quasiparticles are present.
Hence, $\eta_\textrm{b}/\tau$ decreases with $p$ and $T$. On the other hand, the fraction of excess majority fermions increases with $p$, and they increase the shear viscosity. The excess fermions do not participate in pairing and they occupy certain regions in momentum space~\cite{Chienthesis}. In the shallow BEC regime illustrated in Fig.~\ref{fig.2} (b), a competition between a suppression of the non-condensed pairs and an increase of excess-fermions causing gapless excitations as $p$ increases leads to a minimum in the ratio of shear viscosity and relaxation time at intermediate polarization and temperature.

\section{Relation between thermodynamics and transport}\label{sec:relation}
As shown in Ref.~\cite{OurPRA17}, there exists a relation for unpolarized unitary Fermi superfluids connecting thermodynamic quantities, including the pressure and chemical potential, with transport coefficients, including the shear viscosity and superfluid density. The relation is exact at the mean-field level, and an approximate relation was proposed in the presence of pairing fluctuations. Here we derive the analogue relation for homogeneous, polarized unitary Fermi superfluids.

\begin{figure}[th]
	\centering
	\includegraphics[width=0.9\columnwidth, clip]{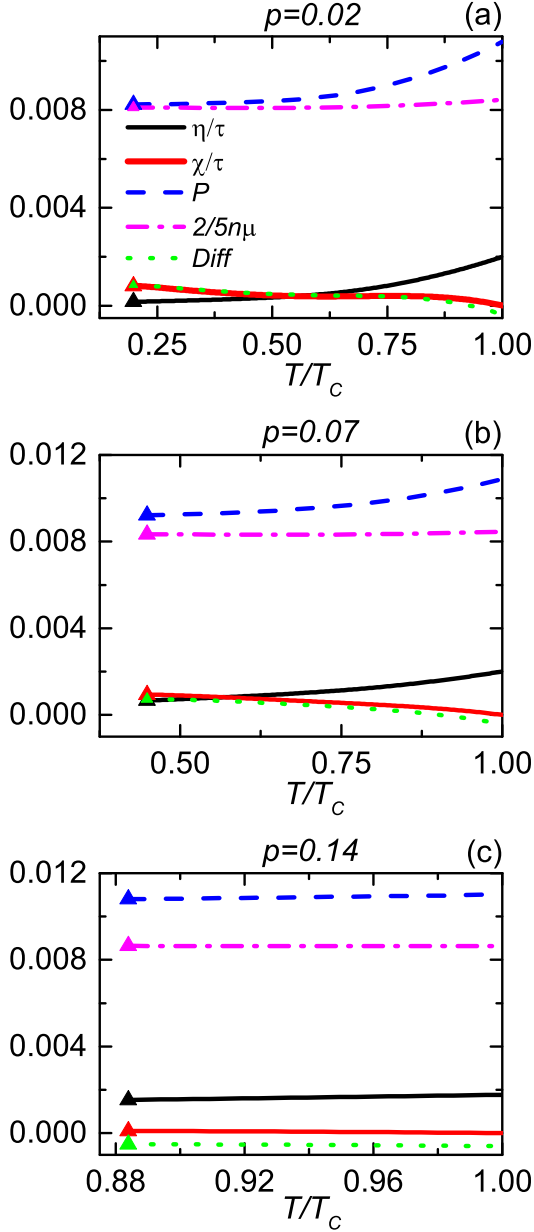}
	\caption{$\eta/\tau$ (black line), $\chi/\tau$ (red line), pressure $P$ (blue dot-dash line), $\frac{2}{5}n\mu$ (pink dashed line) and Diff$\equiv(\eta+\chi)/\tau+\frac{2}{5}n\mu-P$ (green dotted line) of a unitary Fermi gas as a function of temperature (below $T_c$) at $p=$0.02 (a), 0.07 (b) and 0.14 (c). The triangles label where phase separation occurs.}
	\label{fig.pg}
\end{figure}

\subsection{Exact relation at mean-field level}
We start with the mean-field theory and found the following relation
\begin{eqnarray}\label{etaP}
\eta+\chi=(P-\frac{2}{5}\mu n_s)\tau.
\end{eqnarray}
Here $\eta$ is obtained from the mean-field theory with the CFOP theory and its expression is similar to $\eta_\textrm{f}$ except the total gap $\Delta$ plays the role of the order parameter $\Delta_{sc}$, $\chi$ is the anomalous shear viscosity representing the momentum transfer via Cooper pairs, $P$ is the pressure, $n_s$ is the superfluid density, and $\tau$ is the relaxation time. The relation is formally identical to the relation of unpolarized unitary Fermi gases \cite{OurPRA17} except the polarization has been included in all the physical quantities used here.

A derivation of the exact relation at the mean-field level is given in Appendix~\ref{appa1}.
The pressure $P$ is given by
\begin{eqnarray}\label{t0}
P=-\sum_{\mathbf{k}}(\xi_{\mathbf{k}}-E_{\mathbf{k}})-\frac{\Delta^2}{g}+\sum_{\mathbf{k},\sigma}T\ln(1+e^{-\frac{E_{\mathbf{k}\sigma}}{T}}).
\end{eqnarray}
The superfluid density can be obtained from the paramagnetic response function via \cite{Walecka}  \begin{align}n_\textrm{s}=m\lim_{\omega\rightarrow0}\lim_{\mathbf{q}\rightarrow\mathbf{0}}\textrm{Re}[\mathcal{P}^{xx}(\omega,\mathbf{q})]+n.
\end{align} 
For polarized Fermi gases in the BCS-Leggett theory, we found 
\begin{align}\label{nse}
n_{\textrm{s}}=\frac{2\Delta^2}{3m}\sum_{\mathbf{k}}\frac{k^2}{E^2_{\mathbf{k}}}\Big(\frac{1-2\bar{f}(E_{\mathbf{k}})}{2E_{\mathbf{k}}}+\bar{f}'(E_\mathbf{k})\Big),
\end{align}
where $\bar{f}'(x)=\big(f'(x+h)+f'(x-h)\big)/2$.
The shear viscosity characterizes the momentum transfer via the normal density, but the Cooper pairs can also transfer momentum and lead to the anomalous shear viscosity $\chi$ \cite{OurPRA17}. Similar to the stress tensor, we define the anomalous stress tensor $\tensor{\Pi}(\mathbf{x})=\frac{1}{m}\big(\nabla\psi_{\downarrow}(\mathbf{x})\nabla\psi_{\uparrow}(\mathbf{x})+\nabla\psi^\dagger_{\uparrow}(\mathbf{x})\nabla\psi^\dagger_{\downarrow}(\mathbf{x})\big)$. While the shear viscosity is obtained from the stress-stress response function, the anomalous shear viscosity is obtained from the $\tensor{\Pi}$-$\tensor{\Pi}$ response function and is given by
$\chi\equiv-\lim_{\omega\rightarrow0}\lim_{q\rightarrow0}\frac{1}{\omega}\textrm{Im}[Q^{xyxy}(\omega,\mathbf{q})]$. Here $\hat{\tensor{Q}}(\bar{\tau}-\bar{\tau}',\mathbf{q})=-i\theta(\bar{\tau}-\bar{\tau}')\langle[\tensor{\Pi}(\bar{\tau},\mathbf{q}),\tensor{\Pi}(\bar{\tau}',-\mathbf{q})]\rangle$ with $\bar{\tau}$ being the imaginary time and $\theta(x)$ the Heaviside step function.
By incorporating the relaxation time in the same manner as the shear viscosity, we get
\begin{eqnarray}\label{chi}
\chi=-\frac{1}{15}\sum_{\mathbf{k}}\frac{k^4}{m^2}\frac{\Delta^2}{E^2_\mathbf{k}}\Big(\frac{\partial f(E_{\mathbf{k}\uparrow})}{\partial E_{\mathbf{k}\uparrow}}+\frac{\partial f(E_{\mathbf{k}\downarrow})}{\partial E_{\mathbf{k}\downarrow}}\Big)\tau.
\end{eqnarray}
The details are shown in Appendix~\ref{appa1}. Thus, the exact mean-field relation applies to unitary Fermi gases with or without population imbalance. The relation also implies the consistency of the equations of state and linear response theory implemented in this work.

\subsection{Approximate relation beyond mean-field}
In the presence of pairing fluctuations, the exact relation corresponding to Eq.~\eqref{etaP} has not been fully resolved and an approximate relation was proposed instead~\cite{OurPRA17}. Here we follow a similar idea and construct an approximate relation. A natural generalization is to include the contribution from the noncondensed bosons. However, the anomalous shear viscosity only measures the momentum
transfer through the Cooper pairs and we do not include pairing fluctuations there. Instead, an approximate relation for unitary Fermi superfluids with population imbalance is proposed here:
\begin{eqnarray}\label{etaP2}
\eta+\chi\approx(P-\frac{2}{5}\mu n)\tau.
\end{eqnarray}
Here $\eta=\eta_\textrm{f}+\eta_\textrm{b}$, $P=P_\textrm{f}+P_\textrm{b}$ with $P_\textrm{f}$  given by Eq.~(\ref{t0}) and $P_\textrm{b}=-T\sum_\mathbf{q}\ln(1-e^{-\frac{\Omega_\mathbf{q}}{T}})$ being the pressure of noncondensed pairs in the approximation summarized in Appendix~\ref{appb1}, $n$ is the total particle number density, and
\begin{eqnarray}
\chi\approx-\frac{1}{15}\sum_{\mathbf{k}}\frac{k^4}{m^2}\frac{\Delta^2_\textrm{sc}}{E^2_\mathbf{k}}\Big(\frac{\partial f(E_{\mathbf{k}\uparrow})}{\partial E_{\mathbf{k}\uparrow}}+\frac{\partial f(E_{\mathbf{k}\downarrow})}{\partial E_{\mathbf{k}\downarrow}}\Big)\tau
\end{eqnarray}
In our approximation, the anomalous shear viscosity $\chi$ only include the contribution from condensed pairs and it vanishes above $T_c$ since $\Delta_\textrm{sc}(T>T_c)=0$. When $T\rightarrow 0$, this identity reduces to the mean field result (\ref{etaP}).
We use numerical calculations to check the validity of our approximation and present the comparison in Figure~\ref{fig.pg}, where we show $\eta/\tau$, $\chi/\tau$, $P$, and $\frac{2}{5}n\mu$ as a function of $T/T_c$ for $p=0.02, 0.07$, and $0.14$. The corresponding $T_c/T_F$ values are 0.262, 0.252, and 0.198. We caution that it is known the $t$-matrix overestimates $T_c$~\cite{Ourreview}. We also show the deviation from the identity (\ref{etaP2}), called $\textrm{Diff}=(\eta+\chi)/\tau+\frac{2}{5}n\mu-P$. The maximal relative error defined by $|\textrm{Diff}/P|$ is 7.4\% for $p$=0.02, 7.8\% for $p$=0.07 and 5.5\% for $p$=0.14, respectively.
Hence the approximate relation works reasonably.

The relation may help determine the relaxation time $\tau$ if the shear viscosity, pressure, chemical potential, and superfluid density can be measured in Fermi gases. Since the anomalous shear viscosity plays a similar role as the shear viscosity, measurements of the shear viscosity are likely to report the combined value of $\eta$ and $\chi$. Then $\tau$ is the only unknown in the relation and its value can be estimated. Recent progresses on measuring thermodynamic quantities in homogeneous unpolarized~\cite{Horikoshi17} and polarized~\cite{Mukherjee17} Fermi gases may eventually accomplish the task of determining the elusive $\tau$.

\section{Conclusions}\label{sec:conclusion}
The shear viscosity of homogeneous, population-imbalanced Fermi gases in the unitary and BEC regimes has been analyzed because phase separation at low temperatures in the BCS and unitary regimes hinders a full description. The contributions from noncondensed pairs are included by a pairing-fluctuation theory. In general, the ratio between shear viscosity and relaxation time increases with the polarization, but a competition between the noncondensed pairs and excess fermions is found in the shallow BEC regime and it causes a minimum in $\eta/\tau$ as $p$ increases. To help determine the relaxation time and constrain physical quantities, we present a relation for polarized unitary Fermi superfluid connecting the shear viscosity, pressure, superfluid density, chemical potential, and anomalous shear viscosity. Although the relation is exact at the mean-field level, in the presence of pairing fluctuations only an approximation is proposed and its full expression awaits future investigations.

\textit{Acknowledgment}:  H. G. thanks the support from the National
Natural Science Foundation of China (Grant No.
11674051).

\appendix
\section{Shear viscosity from gauge-invariant linear response theory}\label{appeta}
For population-imbalanced Fermi gases, the Hamiltonian respects a global U(1) symmetry $\psi_\sigma\rightarrow e^{-i\alpha}\psi_\sigma$, where $\psi_\sigma$ is the fermionic field for each species.
The current-current response function is evaluated from a gauge invariant linear response theory \cite{OurJPB14}, which can be obtained by ``gauging'' the U(1) symmetry. To implement it, the symmetry becomes a local symmetry and we introduce an effective gauge field to maintain the symmetry. The gauge field, which can be thought of as an effective electromagnetic (EM) field $A^\mu=(\phi,\mathbf{A})$, interacts with the fermionic field by coupling with the Noether current of the U(1) symmetry given by $J^\mu=(n,\mathbf{J})$. Here
\begin{eqnarray}\label{EC}
&
&\mathbf{J}(\mathbf{x})=-\frac{1}{2mi}\sum_\sigma\Big[\psi^{\dagger}_{\sigma}(\mathbf{x})\big(\nabla\psi_{\sigma}(\mathbf{x})\big)-\big(\nabla\psi^{\dagger}_{\sigma}(\mathbf{x})\big)\psi_{\sigma}(\mathbf{x})\Big]\nonumber\\
& &\qquad-\frac{1}{m}\mathbf{A}(\mathbf{x})\sum_\sigma\psi^{\dagger}_{\sigma}(\mathbf{x})\psi_{\sigma}(\mathbf{x}),
\nonumber\\
& &n(\mathbf{x})=\sum_\sigma\psi^{\dagger}_{\sigma}(\mathbf{x})\psi_{\sigma}(\mathbf{x}).
\end{eqnarray}
The conserved current is perturbed by the effective external EM field as $\delta J^{\mu}(Q)=K^{\mu\nu}A_{\nu}(Q)$, where $\delta J^{\mu}$ is the perturbed mass current, and
\begin{eqnarray}\label{Kmn}
& &K^{\mu\nu}(Q)=\frac{n}{m}h^{\mu\nu}\\
&+&\sum_{K\sigma}\Gamma^{\mu}_{\sigma}(K+Q,K)G_{\sigma}(K+Q)\gamma^{\nu}_{\sigma}(K,K+Q)G_{\sigma}(K)\nonumber
\end{eqnarray}
is the EM response function.
Here
$\gamma^{\mu}_{\sigma}(K+Q,K)=S_{\sigma}(1,\frac{\mathbf{p}+\frac{\mathbf{q}}{2}}{m})$ and $S_{\uparrow,\downarrow}=\pm1$ is the bare EM interaction vertex, $\Gamma^{\mu}_{\sigma}(K+Q,K)$ is the full EM interaction vertex and $h^{\mu\nu}=-\eta^{\mu\nu}(1-\eta^{\nu0})$ with $\eta^{\mu\nu}=\textrm{diag}(1,-1,-1,-1)$ being the metric tensor.

In a gauge invariant theory, the vertex must satisfy the Ward identity \cite{Nambu60,Schrieffer_book,OurJLTP13}
\begin{eqnarray}\label{WI0}
q_{\mu}\Gamma^{\mu}_{\sigma}(K+Q,K)=G^{-1}_{\sigma}(K+Q)-G^{-1}_{\sigma}(K).
\end{eqnarray}
It will guarantee that the perturbed current is also conserved: $q_{\mu}\delta J^{\mu}(Q)=0$.
The gauge invariant EM vertex and the response function $\tensor{K}$ for unpolarized Fermi gases within the BCS mean field formalism can be found in Ref.~\cite{OurJPB14}.

The current-current response functions correspond to the spatial part of Eq.~(\ref{Kmn}) and can be decomposed in to the form
\begin{equation}\label{eq:Kdecom}
\tensor{K}=\tensor{\mathcal{P}}+\frac{\tensor{n}}{m}+\tensor{C},
\end{equation}
where $\tensor{n}=n\tensor{1}$ with $\tensor{1}$ being the unit tensor has no imaginary part and gives no contribution to the shear viscosity, and $\tensor{C}$ comes from the contributions of collective modes and does not contribute to the shear viscosity~\cite{OurPRA17}.  Only the paramagnetic response function $\tensor{\mathcal{P}}$ is relevant and its expression is given in Appendix.\ref{appa1}. To obtain the expression of the shear viscosity we follow the formalism in Ref.~\cite{OurPRA17} and incorporate the relaxation time~\cite{Kadanoff61} by regularizing the $\delta$-function with a Lorentzian function
\begin{eqnarray} \label{RG1}
\delta(x)=\lim_{\Gamma\rightarrow0}\frac{1}{\pi}\frac{\Gamma}{x^2+\Gamma^2}.
\end{eqnarray}
Hence the shear viscosity is found to be
\begin{equation}\label{SVm}
\eta=\frac{1}{30\pi^2m^2}\int_0^{\infty}dkk^6\frac{\xi^2_{\mathbf{k}}}{E^2_{\mathbf{k}}}\Big[-\frac{\partial f(E_{\mathbf{k}\uparrow})}{\partial E_{\mathbf{k}\uparrow}}-\frac{\partial f(E_{\mathbf{k}\downarrow})}{\partial E_{\mathbf{k}\downarrow}}\Big]\tau,
\end{equation}
where $\tau=\frac{1}{\Gamma}$ is the relaxation time.

Within the pairing fluctuation formalism consistent with the Leggett-BCS theory~\cite{OurAnnPhys,OurPRA17}, a gauge invariant EM vertex respecting the Ward identity has the form
\begin{eqnarray}\label{V1}
& &\Gamma^{\mu}_{\sigma}(K+Q,K)=\gamma^{\mu}_{\sigma}(K+Q,K)+\Gamma^{\mu}_{\textrm{Coll},\sigma}(K+Q,K) \nonumber \\
&+&\Gamma^{\mu}_{\textrm{MT},\textrm{sc},\sigma}(K+Q,K)+\Gamma^{\mu}_{\textrm{MT},\textrm{pg},\sigma}(K+Q,K)\nonumber\\
&+&\Gamma^{\mu}_{\textrm{AL},1,\sigma}(K+Q,K)+\Gamma^{\mu}_{\textrm{AL},2,\sigma}(K+Q,K).
\end{eqnarray}
The second term $\Gamma^{\mu}_{\textrm{Coll},\sigma}(K+Q,K)$ in the expression stands for the contributions from the collective modes due to the spontaneous breaking of the U(1) symmetry in the superfluid phase.
However, this term is irrelevant when we derive the shear viscosity~\cite{OurJLTP13}, so we skip it full expression. The third and fourth terms come from the Maki-Thompson (MT) diagrams associated with the condensed and non-condensed pairs, respectively, and the fifth and sixth terms are two Aslamazov-Larkin (AL) diagrams introduced in a way satisfying the Ward identity. The expressions of those diagrams can be found in Appendix~\ref{appb1}.  By using the identity (\ref{ALMT}), the paramagnetic response function is given by Eq.~(\ref{D-Qij}). It can be proven that this formalism satisfies the sum rule~\cite{KM2,HaoNJP11}
\begin{eqnarray}
\lim_{\mathbf{q}\rightarrow0}\int^{\infty}_{\infty}\Big(-\frac{\textrm{Im}K_{\textrm{T}}(\omega,\mathbf{q})}{\omega}\Big)=\frac{n_{\textrm{n}}(T)}{m}.
\end{eqnarray}
Here $n_{\textrm{n}}(T)=n-n_s(T)$ is the normal-fluid density.
All the expressions apply to population-imbalanced Fermi gases when the corresponding thermodynamic quantities are used.

\section{Details for Mean-Field Theory}\label{appa1}
At the mean-field level, we define
$E^{\pm}_{\mathbf{k}}=E_{\mathbf{k}\pm\frac{\mathbf{q}}{2}}$, $E^{\pm}_{\mathbf{k}\downarrow,\uparrow}=E^{\pm}_{\mathbf{k}}\mp h$ and let $K^{\mu\nu}=Q^{\mu\nu}+\frac{n}{m}h^{\mu\nu}$. The paramagnetic current-current response function can be derived from Eq.~\eqref{eq:Kdecom} and is given by
\begin{widetext}
\begin{align}\label{Pij}
\tensor{\mathcal{P}}^{ij}(\omega,\mathbf{q})=\sum_{\mathbf{k}}\frac{\mathbf{k}^i\mathbf{k}^j}{2m^2}\Big\{\Big(1-\frac{\xi^+_{\mathbf{k}}\xi^-_{\mathbf{k}}+\Delta^2}{E^+_{\mathbf{k}}E^-_{\mathbf{k}}}\Big)
&\Big(\frac{1-f(E^+_{\mathbf{k}\uparrow})-f(E^-_{\mathbf{k}\downarrow})}{\omega-E^+_{\mathbf{k}\uparrow}-E^-_{\mathbf{k}\downarrow}}-\frac{1-f(E^+_{\mathbf{k}\downarrow})-f(E^-_{\mathbf{k}\uparrow})}{\omega+E^+_{\mathbf{k}\downarrow}+E^-_{\mathbf{k}\uparrow}}\Big)
\notag\\
-\Big(1+\frac{\xi^+_{\mathbf{k}}\xi^-_{\mathbf{k}}+\Delta^2}{E^+_{\mathbf{k}}E^-_{\mathbf{k}}}\Big)
&\Big(\frac{f(E^+_{\mathbf{k}\uparrow})-f(E^-_{\mathbf{k}\uparrow})}{\omega-E^+_{\mathbf{k}\uparrow}+E^-_{\mathbf{k}\uparrow}}-\frac{f(E^+_{\mathbf{k}\downarrow})-f(E^-_{\mathbf{k}\downarrow})}{\omega+E^+_{\mathbf{k}\downarrow}-E^-_{\mathbf{k}\downarrow}}\Big)\Big\}.
\end{align}
\end{widetext}
The expression (\ref{SVm}) of the shear viscosity can be derived by similar steps leading to  Eq.~(\ref{DSV1}).

Now we derive an expression of the anomalous shear viscosity following Ref.~\cite{OurPRA17}. The interaction vertex has a dyadic form in the Nambu space $\tensor{\gamma}(K,K+Q)=\frac{\mathbf{k}(\mathbf{k}+\mathbf{q})}{m^2}\sigma_1$ with $\sigma_1$ being the first Pauli matrix. Here we present the derivation of the $\tensor{\Pi}-\tensor{\Pi}$ correlation function in the presence of population imbalance. After applying the Fourier transform and using Wick's theorem, we get
\begin{align}
& \hat{\tensor{Q}}(i\Omega_l,\mathbf{q})\notag\\&=T\sum_{i\omega_n}\sum_{\mathbf{k}}\textrm{Tr}\big(\tensor{\gamma}(K,K+Q)\hat{G}(K+Q)\tensor{\gamma}(K+Q,K)\hat{G}(K)\big)\notag\\
&=T\sum_{i\omega_n}\sum_{\mathbf{k}}\frac{\mathbf{k}(\mathbf{k}+\mathbf{q})}{m^2}\frac{(\mathbf{k}+\mathbf{q})\mathbf{k}}{m^2}\textrm{Tr}\big(\sigma_1\hat{G}(K+Q)\sigma_1\hat{G}(K)\big)\notag\\
&=T\sum_{i\omega_n}\sum_{\mathbf{k}}\frac{\mathbf{k}(\mathbf{k}+\mathbf{q})(\mathbf{k}+\mathbf{q})\mathbf{k}}{m^4}\big(2F_{\uparrow\downarrow}(K+Q)F_{\uparrow\downarrow}(K)\notag\\
&-G_\downarrow(-K-Q)G_\uparrow(K)-G_\uparrow(K+Q)G_\downarrow(-K)\big).
\end{align}
Here $\hat{G}(K)=\left(\begin{array}{ll} G_{\uparrow}(K) & F_{\uparrow\downarrow}(K) \\ F_{\downarrow\uparrow}(-K) & -G_{\downarrow}(-K)\end{array}\right)$ is the Green's function in the Nambu space~\cite{OurJPB14} and
\begin{align}F_{\sigma\bar{\sigma}}(K)=-\frac{\Delta}{(i\omega_n-E_{\mathbf{k}\sigma})(i\omega_n+E_{\mathbf{k}\bar{\sigma}})}\end{align}
is the anomalous Green's function. It has the property $F_{\uparrow\downarrow}(-K)=F_{\downarrow\uparrow}(K)$.  After plugging in the expressions of Green's functions and following a complex continuation, we get
\begin{widetext}\begin{align}
\hat{\tensor{Q}}(\omega,\mathbf{q})=\sum_{\mathbf{k}}\frac{\mathbf{k}^-\mathbf{k}^+\mathbf{k}^+\mathbf{k}^-}{2m^4}\Big\{&\big(1+\frac{\xi^+_{\mathbf{k}}\xi^-_{\mathbf{k}}-\Delta^2}{E^+_{\mathbf{k}}E^-_{\mathbf{k}}}\big)\Big(\frac{1-f(E^+_{\mathbf{k}\uparrow})-f(E^-_{\mathbf{k}\downarrow})}{\omega-E^+_{\mathbf{k}\uparrow}-E^-_{\mathbf{k}\downarrow}}-\frac{1-f(E^+_{\mathbf{k}\downarrow})-f(E^-_{\mathbf{k}\uparrow})}{\omega+E^+_{\mathbf{k}\downarrow}+E^-_{\mathbf{k}\uparrow}}\Big) \notag \\
-&\big(1-\frac{\xi^+_{\mathbf{k}}\xi^-_{\mathbf{k}}-\Delta^2}{E^+_{\mathbf{k}}E^-_{\mathbf{k}}}\big)\Big(\frac{f(E^+_{\mathbf{k}\uparrow})-f(E^-_{\mathbf{k}\uparrow})}{\omega-E^+_{\mathbf{k}\uparrow}+E^-_{\mathbf{k}\uparrow}}-\frac{f(E^+_{\mathbf{k}\downarrow})-f(E^-_{\mathbf{k}\downarrow})}{\omega+E^+_{\mathbf{k}\downarrow}-E^-_{\mathbf{k}\downarrow}}\Big)\Big\},
\end{align}
\end{widetext}
where $\mathbf{k}^\pm=\mathbf{k}\pm\frac{\mathbf{q}}{2}$. By following the same step of Eq.~(\ref{DSV1}) to incorporate the relaxation time, the anomalous shear viscosity is
\begin{align}
\chi
=-\frac{1}{15}\sum_{\mathbf{k}}\frac{k^4}{m^2}\frac{\Delta^2}{E^2_\mathbf{k}}\Big(\frac{\partial f(E_{\mathbf{k}\uparrow})}{\partial E_{\mathbf{k}\uparrow}}+\frac{\partial f(E_{\mathbf{k}\downarrow})}{\partial E_{\mathbf{k}\downarrow}}\Big)\tau.
\end{align}

Now we are ready to give a brief proof of the relation (\ref{etaP}). We first prove $P=\frac{2}{3}E$ for polarized unitary Fermi gas in the superfluid phase, where
\begin{eqnarray}\label{te}
E=\sum_{\mathbf{k}}(\xi_{\mathbf{k}}-E_{\mathbf{k}})+\frac{\Delta^2}{g}+2\sum_{\mathbf{k}}E_{\mathbf{k}}\bar{f}(E_{\mathbf{k}})+\mu n
\end{eqnarray}
is the energy density. Integrating by parts, we get
\begin{align}\label{tmp1}
& \sum_{\mathbf{k},\sigma}T\ln(1+e^{-\frac{E_{\mathbf{k}\sigma}}{T}})=\frac{2}{3}\sum_{\mathbf{k}}\frac{k^2}{m}\frac{\xi_{\mathbf{k}}}{E_{\mathbf{k}}}\bar{f}(E_{\mathbf{k}}),\notag\\
& \sum_\mathbf{k}(\xi_\mathbf{k}-E_\mathbf{k}+\frac{\Delta^2}{2\epsilon_\mathbf{k}})=-\frac{1}{3m}\sum_\mathbf{k} k^2\left(1-\frac{\xi_\mathbf{k}}{E_\mathbf{k}}-\frac{\Delta^2}{2\epsilon^2_\mathbf{k}}\right).
\end{align}
Substituting these identities to the expressions of the pressure and energy, and using $\frac{1}{g}=\sum_\mathbf{k}\frac{1}{2\epsilon_\mathbf{k}}$ in the unitary limit, we obtain
\begin{align}
E-\frac{3}{2}P=\Delta^2\sum_\mathbf{k}\left(\frac{1}{\epsilon_\mathbf{k}}-\frac{1}{E_{\mathbf{k}}}+\frac{2}{E_{\mathbf{k}}}\bar{f}(E_{\mathbf{k}})\right)=0,
\end{align}
where the gap equation has been applied.

Now we prove the relation (\ref{etaP}). Integrating by parts and applying Eq.~(\ref{tmp1}), the expression of $\eta$ becomes
\begin{eqnarray}\label{eta2}
\eta&=&P\tau+\sum_{\mathbf{k}}(\xi_{\mathbf{k}}-E_{\mathbf{k}}+\frac{\Delta^2}{2\epsilon_{\mathbf{k}}})\tau\nonumber\\
&+&\frac{1}{15\pi^2m}\int_{0}^{+\infty}dk\frac{k^6}{m}\frac{\Delta^2}{E^3_{\mathbf{k}}}\bar{f}(E_{\mathbf{k}})\tau.
\end{eqnarray}
By applying $E=\frac{3}{2}P$ and Eq.~(\ref{tmp1}), we have
\begin{eqnarray}
& &\sum_{\mathbf{k}}(\xi_{\mathbf{k}}-E_{\mathbf{k}}+\frac{\Delta^2}{2\epsilon_{\mathbf{k}}})\\
&=&-\frac{2}{5}\Big(-3\sum_{\mathbf{k}}T\ln(1+e^{-\frac{E_{\mathbf{k}}}{T}})+2\sum_{\mathbf{k}}E_{\mathbf{k}}\bar{f}(E_{\mathbf{k}})+\mu n\Big)\nonumber\\
&=&-\frac{2}{5}\sum_{\mathbf{k}}\Big[2\frac{\Delta^2}{E_{\mathbf{k}}}\bar{f}(E_{\mathbf{k}})+\mu\big(1-
\frac{\xi_{\mathbf{k}}}{E_{\mathbf{k}}}\big)\Big]\nonumber\\
&=&-\frac{1}{15\pi^2m}\int_{0}^{+\infty}dk\frac{k^6}{m}\frac{\Delta^2}{E^3_{\mathbf{k}}}\bar{f}(E_{\mathbf{k}})\nonumber\\
&-&\frac{1}{15\pi^2}\int_{0}^{+\infty}dk\frac{\Delta^2}{E^2_{\mathbf{k}}}\frac{k^4}{m}\Big[\mu\frac{1-2
\bar{f}(E_{\mathbf{k}})}{E_{\mathbf{k}}}-2\xi_{\mathbf{k}}\bar{f}'(E_{\mathbf{k}})\Big].\nonumber
\end{eqnarray}
After substituting this result into Eq.~(\ref{eta2}), we finally get
\begin{eqnarray}\label{etaP20}
\eta&=&P\tau\nonumber\\&-&\frac{2}{15\pi^2}\int_{0}^{+\infty}dk\frac{\Delta^2}{E^2_{\mathbf{k}}}\frac{k^4}{m}\Big[\mu\frac{1-2
\bar{f}(E_{\mathbf{k}})}{2E_{\mathbf{k}}}-\xi_{\mathbf{k}}\bar{f}'(E_{\mathbf{k}})\Big]\tau\nonumber\\
&=&P\tau-\frac{2}{5}\mu n_{\textrm{s}}\tau+\frac{2}{15}\sum_{\mathbf{k}}\frac{\Delta^2}{E^2_{\mathbf{k}}}\frac{k^4}{m^2}\bar{f}'(E_{\mathbf{k}})\tau\nonumber\\
&=&(P-\frac{2}{5}\mu n_{\textrm{s}})\tau-\chi,
\end{eqnarray}
where we have used the expressions (\ref{nse}) and (\ref{chi}).

\section{Details for Pairing Fluctuation Theory}\label{appb1}
The MT and AL diagrams for obtaining the gauge-invariant vertex are given as follows.
\begin{eqnarray}
& &\Gamma^{\mu}_{\textrm{MT},\textrm{sc},\sigma}(K+Q,K)=\sum_Lt_{\textrm{sc}}(L)G_{0\bar{\sigma}}(L-K)\nonumber\\
&\times&\gamma^{\mu}_{\bar{\sigma}}(L-K,L-K-Q)G_{0\bar{\sigma}}(L-K-Q),
\end{eqnarray}
\begin{eqnarray}
& &\Gamma^{\mu}_{\textrm{MT},\textrm{pg},\sigma}(K+Q,K)=\sum_Lt_{\textrm{pg}}(L)G_{0\bar{\sigma}}(L-K)\nonumber\\
&\times&\gamma^{\mu}_{\bar{\sigma}}(L-K,L-K-Q)G_{0\bar{\sigma}}(L-K-Q),
\end{eqnarray}
\begin{eqnarray}
& &\Gamma^{\mu}_{\textrm{AL},1,\sigma}(K+Q,K)=-\sum_{L,M}t_{\textrm{pg}}(L)t_{\textrm{pg}}(L+Q)\nonumber\\
&\times&G_{0\bar{\sigma}}(L-K)G_{\sigma}(L-M)G_{0\bar{\sigma}}(M+Q)\nonumber\\
&\times&\gamma^{\mu}_{\bar{\sigma}}(M+Q,M)G_{0\bar{\sigma}}(M),
\end{eqnarray}
and
\begin{eqnarray}
& &\Gamma^{\mu}_{\textrm{AL},2,\sigma}(P+Q,P)=-\sum_{L,M}t_{\textrm{pg}}(L)t_{\textrm{pg}}(L+Q)\nonumber\\
&\times&G_{0\bar{\sigma}}(L-K)G_{0\bar{\sigma}}(L-M)G_{\sigma}(M+Q)\nonumber\\
&\times&\Gamma^{\mu}_{\sigma}(M+Q,M)G_{\sigma}(M),
\end{eqnarray}
where $t_{\textrm{sc}}$ and $t_{\textrm{pg}}$ are the $t$-matrices associated with the condensed and non-condensed pairs, respectively.
Moreover, the AL and MT$_{\textrm{pg}}$ diagrams satisfy an identity 
\begin{eqnarray}\label{ALMT}
& &q_{\mu}\big[\frac{1}{2}\Gamma^{\mu}_{\textrm{AL},1,\sigma}(K+Q,K)+\frac{1}{2}\Gamma^{\mu}_{\textrm{AL},2,\sigma}(K+Q,K)\nonumber\\&+&\Gamma^{\mu}_{\textrm{MT},\textrm{pg},\sigma}(K+Q,K)\big]=0,
\end{eqnarray}
which brings further simplification to our evaluation of the shear viscosity.

Including the pairing fluctuation effects, the expression of the paramagnetic response function is given by
\begin{widetext}
\begin{align}\label{D-Qij}
\tensor{\mathcal{P}}^{ij}(\omega,\mathbf{q})=\sum_{\mathbf{k}}\frac{\mathbf{k}^i\mathbf{k}^j}{2m^2}\Big\{&\Big(1-\frac{\xi^+_{\mathbf{k}}\xi^-_{\mathbf{k}}+\Delta^2_\textrm{sc}-\Delta^2_\textrm{pg}}{E^+_{\mathbf{k}}E^-_{\mathbf{k}}}\Big)
\Big(\frac{1-f(E^+_{\mathbf{k}\uparrow})-f(E^-_{\mathbf{k}\downarrow})}{\omega-E^+_{\mathbf{k}\uparrow}-E^-_{\mathbf{k}\downarrow}}-\frac{1-f(E^+_{\mathbf{k}\downarrow})-f(E^-_{\mathbf{k}\uparrow})}{\omega+E^+_{\mathbf{k}\downarrow}+E^-_{\mathbf{k}\uparrow}}\Big)
\notag\\
-&\Big(1+\frac{\xi^+_{\mathbf{k}}\xi^-_{\mathbf{k}}+\Delta^2_\textrm{sc}-\Delta^2_\textrm{pg}}{E^+_{\mathbf{k}}E^-_{\mathbf{k}}}\Big)
\Big(\frac{f(E^+_{\mathbf{k}\uparrow})-f(E^-_{\mathbf{k}\uparrow})}{\omega-E^+_{\mathbf{k}\uparrow}+E^-_{\mathbf{k}\uparrow}}-\frac{f(E^+_{\mathbf{k}\downarrow})-f(E^-_{\mathbf{k}\downarrow})}{\omega+E^+_{\mathbf{k}\downarrow}-E^-_{\mathbf{k}\downarrow}}\Big)\Big\}.
\end{align}
\end{widetext}
In the mean-field BCS-Leggett theory, $\Delta_\textrm{pg}=0$ and $\Delta_\textrm{sc}=\Delta$, and this expression reduces to Eq.~(\ref{Pij}).

The shear viscosity acquire two contributions, $\eta=\eta_\textrm{f}+\eta_\textrm{b}$.
The fermionic contribution to the shear viscosity, including the fermionic quasiparticles and condensed pairs, is given by Eq.~(\ref{eta0}).
 In the limit $q\rightarrow0$, we have
\begin{eqnarray}
& &E^+_{\mathbf{k}\sigma}-E^-_{\mathbf{k}\sigma}
=E^+_{\mathbf{k}}-E^-_{\mathbf{k}}\nonumber\\&=&\mathbf{q}\cdot\nabla E_{\mathbf{k}}=\frac{\mathbf{k}\cdot\mathbf{q}}{m}\frac{\xi_{\mathbf{k}}}{E_{\mathbf{k}}}=\frac{pq\cos\theta}{m}\frac{\xi_{\mathbf{k}}}{E_{\mathbf{k}}}.
\end{eqnarray}
To derive the expression of the shear viscosity, we need to regularize the $\delta$-function coming from the imaginary part of the response function given by Eq.(\ref{RG1}),  \begin{eqnarray}\delta(\omega\pm\mathbf{q}\cdot\nabla_{\mathbf{k}} E)=
\lim_{\Gamma \rightarrow0}
\frac{\frac{1}{\pi}\Gamma}{(\omega\pm\mathbf{q}\cdot\nabla E_{\mathbf{k}})^2+
\Gamma^2}.\end{eqnarray}
Hence the shear viscosity is evaluated as
\begin{widetext}
\begin{eqnarray}\label{DSV1}
& &\eta_\textrm{f}=-m^2\lim_{\omega\rightarrow0}\lim_{q\rightarrow0}\frac{\pi\omega}{2q^2}\sum_{\mathbf{k}}\frac{k^2\textrm{sin}^2\theta}{m^2}\Big[\frac{E^+_{\mathbf{k}}E^-_{\mathbf{k}}-\xi^+_{\mathbf{k}}\xi^-_{\mathbf{k}}-\Delta^2_{\textrm{sc}}+\Delta^2_{\textrm{pg}}}{2E^+_{\mathbf{k}}E^-_{\mathbf{k}}}\nonumber\\
&\times&
\big(1-f(E^+_{\mathbf{k}\downarrow})-f(E^-_{\mathbf{k}\uparrow})\big)\delta(\omega+E^+_{\mathbf{k}\downarrow}+E^-_{\mathbf{k}\uparrow})-\big(1-f(E^+_{\mathbf{k}\uparrow})-f(E^-_{\mathbf{k}\downarrow})\big)\delta(\omega-E^+_{\mathbf{k}\uparrow}-E^-_{\mathbf{k}\downarrow})\big)\nonumber\\
&-&\frac{E^+_{\mathbf{k}}E^-_{\mathbf{k}}+\xi^+_{\mathbf{k}}\xi^-_{\mathbf{k}}+\Delta^2_{\textrm{sc}}-\Delta^2_{\textrm{pg}}}{2E^+_{\mathbf{k}}E^-_{\mathbf{k}}}\big(f(E^+_{\mathbf{k}\downarrow})-f(E^-_{\mathbf{k}\downarrow})\big)\delta(\omega+E^+_{\mathbf{k}\downarrow}-E^-_{\mathbf{k}\downarrow})-\big(f(E^+_{\mathbf{k}\uparrow})-f(E^-_{\mathbf{k}\uparrow})\big)
\delta(\omega-E^+_{\mathbf{k}\uparrow}+E^-_{\mathbf{k}\uparrow})\big)\Big]\nonumber\\
&=&-\frac{1}{30\pi m^2}\int_0^{\infty}dkk^6\Big(1-\frac{\Delta^2_{\textrm{pg}}}{E^2_{\mathbf{k}}}\Big)\frac{\xi^2_{\mathbf{k}}}{E^2_{\mathbf{k}}}\lim_{\omega\rightarrow0}\lim_{q\rightarrow0}\big(\frac{\partial f(E_{\mathbf{k}\downarrow})}{\partial E_{\mathbf{k}\downarrow}}\delta(\omega+\mathbf{q}\cdot\nabla E_{\mathbf{k}})+\frac{\partial f(E_{\mathbf{k}\uparrow})}{\partial E_{\mathbf{k}\uparrow}}\delta(\omega-\mathbf{q}\cdot\nabla E_{\mathbf{k}})\big)\nonumber\\
&=&\frac{1}{30\pi^2m^2}\int_0^{\infty}dkk^6\Big(1-\frac{\Delta^2_{\textrm{pg}}}{E^2_{\mathbf{k}}}\Big)\frac{\xi^2_{\mathbf{k}}}{E^2_{\mathbf{k}}}\Big[-\frac{\partial f(E_{\mathbf{k}\uparrow})}{\partial E_{\mathbf{k}\uparrow}}-\frac{\partial f(E_{\mathbf{k}\downarrow})}{\partial E_{\mathbf{k}\downarrow}}\Big]\tau.
\end{eqnarray}
\end{widetext}
Note the $\delta$-functions in the second line vanish because  $E^+_{\mathbf{k}\sigma}+E^-_{\mathbf{k}\bar{\sigma}}=E^+_{\mathbf{k}}+E^-_{\mathbf{k}}>2\Delta$ but $\omega\rightarrow0$ so the argument does not vanish.

The bosonic contribution is from the noncondensed pairs, and it can be obtained by considering the shear viscosity of a gas of composite bosons with the Hamiltonian $H_\textrm{b}=\sum_\mathbf{q}\Omega_\mathbf{q}b^\dagger_\mathbf{q}b_\mathbf{q}$. Here $b_\mathbf{q}$ is the effective annilation operator for the composite bosons. The bosonic Green's function is then given by
\begin{align}
G_\textrm{b}(i\Omega_l,\mathbf{q}) =\frac{1}{i\Omega_l-\Omega_\mathbf{q}}.
\end{align}
The current operator is
\begin{align}
\mathbf{J}_{\textrm{b}}(\bar{\tau},\mathbf{q})=-\frac{1}{M^*}\sum_\mathbf{k}(\mathbf{k}+\frac{\mathbf{q}}{2})b^\dagger_\mathbf{k}(\bar{\tau})b_{\mathbf{k}+\mathbf{q}}(\bar{\tau}).
\end{align}
This defines a $\mathbf{J}-\mathbf{J}$ linear response, and the response function is given by
\begin{equation}
\tensor{Q}_\textrm{b}^{\mathbf{J}\mathbf{J}}(\bar{\tau}-\bar{\tau}',\mathbf{q})=-i\theta(\bar{\tau}-\bar{\tau}')\langle[\mathbf{J}_{\textrm{b}}(\bar{\tau},\mathbf{q}),\mathbf{J}_{\textrm{b}}(\bar{\tau}',-\mathbf{q})]\rangle.
\end{equation}
Finally, the shear viscosity from the noncondensed pairs is given by
\begin{align}
\eta_\textrm{b}&=-M^{\ast2}\lim_{\omega\rightarrow0}\lim_{q\rightarrow0}\textrm{Im}\frac{\omega}{q^2}Q^{\mathbf{J}\mathbf{J}}_{\textrm{b}\textrm{T}}(\omega,\mathbf{q})\notag\\
&=-\frac{1}{30\pi^2M^{\ast2}}\int_0^\infty dkk^6\frac{\partial b(\Omega_\mathbf{k})}{\partial \Omega_\mathbf{k}}\tau.
\end{align}

\bibliographystyle{apsrev}

\begin{thebibliography}{59}
	\expandafter\ifx\csname natexlab\endcsname\relax\def\natexlab#1{#1}\fi
	\expandafter\ifx\csname bibnamefont\endcsname\relax
	\def\bibnamefont#1{#1}\fi
	\expandafter\ifx\csname bibfnamefont\endcsname\relax
	\def\bibfnamefont#1{#1}\fi
	\expandafter\ifx\csname citenamefont\endcsname\relax
	\def\citenamefont#1{#1}\fi
	\expandafter\ifx\csname url\endcsname\relax
	\def\url#1{\texttt{#1}}\fi
	\expandafter\ifx\csname urlprefix\endcsname\relax\def\urlprefix{URL }\fi
	\providecommand{\bibinfo}[2]{#2}
	\providecommand{\eprint}[2][]{\url{#2}}
	
	\bibitem[{\citenamefont{Kovtun et~al.}(2005{\natexlab{a}})\citenamefont{Kovtun,
			Son, and Starinets}}]{Son1}
	\bibinfo{author}{\bibfnamefont{P.~K.} \bibnamefont{Kovtun}},
	\bibinfo{author}{\bibfnamefont{D.~T.} \bibnamefont{Son}}, \bibnamefont{and}
	\bibinfo{author}{\bibfnamefont{A.~O.} \bibnamefont{Starinets}},
	\bibinfo{journal}{Phys. Rev. Lett.} \textbf{\bibinfo{volume}{94}},
	\bibinfo{pages}{111601} (\bibinfo{year}{2005}{\natexlab{a}}).
	
	\bibitem[{\citenamefont{Kinast et~al.}(2005)\citenamefont{Kinast, Turlapov, and
			Thomas}}]{KinastPRL05}
	\bibinfo{author}{\bibfnamefont{J.}~\bibnamefont{Kinast}},
	\bibinfo{author}{\bibfnamefont{A.}~\bibnamefont{Turlapov}}, \bibnamefont{and}
	\bibinfo{author}{\bibfnamefont{J.~E.} \bibnamefont{Thomas}},
	\bibinfo{journal}{Phys. Rev. Lett.} \textbf{\bibinfo{volume}{94}},
	\bibinfo{pages}{170404} (\bibinfo{year}{2005}).
	
	\bibitem[{\citenamefont{Bruun and Smith}(2007)}]{BruunPRA07}
	\bibinfo{author}{\bibfnamefont{G.~M.} \bibnamefont{Bruun}} \bibnamefont{and}
	\bibinfo{author}{\bibfnamefont{H.}~\bibnamefont{Smith}},
	\bibinfo{journal}{Phys. Rev. A} \textbf{\bibinfo{volume}{75}},
	\bibinfo{pages}{043612} (\bibinfo{year}{2007}).
	
	\bibitem[{\citenamefont{Schafer}(2007)}]{SchaferPRA07}
	\bibinfo{author}{\bibfnamefont{T.}~\bibnamefont{Schafer}},
	\bibinfo{journal}{Phys. Rev. A} \textbf{\bibinfo{volume}{76}},
	\bibinfo{pages}{063618} (\bibinfo{year}{2007}).
	
	\bibitem[{\citenamefont{He et~al.}(2007)\citenamefont{He, Chien, Chen, and
			Levin}}]{ourlongpaper}
	\bibinfo{author}{\bibfnamefont{Y.}~\bibnamefont{He}},
	\bibinfo{author}{\bibfnamefont{C.~C.} \bibnamefont{Chien}},
	\bibinfo{author}{\bibfnamefont{Q.~J.} \bibnamefont{Chen}}, \bibnamefont{and}
	\bibinfo{author}{\bibfnamefont{K.}~\bibnamefont{Levin}},
	\bibinfo{journal}{Phys. Rev. B} \textbf{\bibinfo{volume}{76}},
	\bibinfo{pages}{224516} (\bibinfo{year}{2007}).
	
	\bibitem[{\citenamefont{Kinast et~al.}(2008)\citenamefont{Kinast, Turlapov,
			Clancy, Luo, Joseph, and Thomas}}]{ThomasJLTP08}
	\bibinfo{author}{\bibfnamefont{J.}~\bibnamefont{Kinast}},
	\bibinfo{author}{\bibfnamefont{A.}~\bibnamefont{Turlapov}},
	\bibinfo{author}{\bibfnamefont{B.}~\bibnamefont{Clancy}},
	\bibinfo{author}{\bibfnamefont{L.}~\bibnamefont{Luo}},
	\bibinfo{author}{\bibfnamefont{J.}~\bibnamefont{Joseph}}, \bibnamefont{and}
	\bibinfo{author}{\bibfnamefont{J.~E.} \bibnamefont{Thomas}},
	\bibinfo{journal}{J. Low Temp. Phys.} \textbf{\bibinfo{volume}{150}},
	\bibinfo{pages}{567} (\bibinfo{year}{2008}).
	
	\bibitem[{\citenamefont{Nascimbene et~al.}(2010)\citenamefont{Nascimbene,
			Navon, Jian, Chevy, and Salomon}}]{Nascimbene10}
	\bibinfo{author}{\bibfnamefont{S.}~\bibnamefont{Nascimbene}},
	\bibinfo{author}{\bibfnamefont{N.}~\bibnamefont{Navon}},
	\bibinfo{author}{\bibfnamefont{K.~J.} \bibnamefont{Jian}},
	\bibinfo{author}{\bibfnamefont{F.}~\bibnamefont{Chevy}}, \bibnamefont{and}
	\bibinfo{author}{\bibfnamefont{C.}~\bibnamefont{Salomon}},
	\bibinfo{journal}{Nature} \textbf{\bibinfo{volume}{463}},
	\bibinfo{pages}{1057} (\bibinfo{year}{2010}).
	
	\bibitem[{\citenamefont{Enss}(2012)}]{EnssPRA12}
	\bibinfo{author}{\bibfnamefont{T.}~\bibnamefont{Enss}}, \bibinfo{journal}{Phys.
		Rev. A} \textbf{\bibinfo{volume}{86}}, \bibinfo{pages}{013616}
	(\bibinfo{year}{2012}).
	
	\bibitem[{\citenamefont{Elliott et~al.}(2014)\citenamefont{Elliott, Joseph, and
			Thomas}}]{ThomasPRL14}
	\bibinfo{author}{\bibfnamefont{E.}~\bibnamefont{Elliott}},
	\bibinfo{author}{\bibfnamefont{J.~A.} \bibnamefont{Joseph}},
	\bibnamefont{and} \bibinfo{author}{\bibfnamefont{J.~E.}
		\bibnamefont{Thomas}}, \bibinfo{journal}{Phys. Rev. Lett.}
	\textbf{\bibinfo{volume}{113}}, \bibinfo{pages}{020406}
	(\bibinfo{year}{2014}).
	
	\bibitem[{\citenamefont{Cao et~al.}(2011{\natexlab{a}})\citenamefont{Cao,
			Elliott, Joseph, Wu, Petricka, Schafer, and Thomas}}]{ThomasScience11}
	\bibinfo{author}{\bibfnamefont{C.}~\bibnamefont{Cao}},
	\bibinfo{author}{\bibfnamefont{E.}~\bibnamefont{Elliott}},
	\bibinfo{author}{\bibfnamefont{J.}~\bibnamefont{Joseph}},
	\bibinfo{author}{\bibfnamefont{H.}~\bibnamefont{Wu}},
	\bibinfo{author}{\bibfnamefont{J.}~\bibnamefont{Petricka}},
	\bibinfo{author}{\bibfnamefont{T.}~\bibnamefont{Schafer}}, \bibnamefont{and}
	\bibinfo{author}{\bibfnamefont{J.~E.} \bibnamefont{Thomas}},
	\bibinfo{journal}{Science} \textbf{\bibinfo{volume}{472}},
	\bibinfo{pages}{201} (\bibinfo{year}{2011}{\natexlab{a}}).
	
	\bibitem[{\citenamefont{Bruun and Pethick}(2011)}]{PethickPRL11}
	\bibinfo{author}{\bibfnamefont{G.~M.} \bibnamefont{Bruun}} \bibnamefont{and}
	\bibinfo{author}{\bibfnamefont{C.~J.} \bibnamefont{Pethick}},
	\bibinfo{journal}{Phys. Rev. Lett.} \textbf{\bibinfo{volume}{107}},
	\bibinfo{pages}{255302} (\bibinfo{year}{2011}).
	
	\bibitem[{\citenamefont{Sommer et~al.}(2011)\citenamefont{Sommer, Ku, Roati,
			and Zwierlein}}]{ZwierleinNature11}
	\bibinfo{author}{\bibfnamefont{A.}~\bibnamefont{Sommer}},
	\bibinfo{author}{\bibfnamefont{M.}~\bibnamefont{Ku}},
	\bibinfo{author}{\bibfnamefont{G.}~\bibnamefont{Roati}}, \bibnamefont{and}
	\bibinfo{author}{\bibfnamefont{M.~W.} \bibnamefont{Zwierlein}},
	\bibinfo{journal}{Nature} \textbf{\bibinfo{volume}{472}},
	\bibinfo{pages}{201} (\bibinfo{year}{2011}).
	
	\bibitem[{\citenamefont{Wlazlowski et~al.}(2013)\citenamefont{Wlazlowski,
			Magierski, Bulgac, and Roche}}]{ProchePRA13}
	\bibinfo{author}{\bibfnamefont{G.}~\bibnamefont{Wlazlowski}},
	\bibinfo{author}{\bibfnamefont{P.}~\bibnamefont{Magierski}},
	\bibinfo{author}{\bibfnamefont{A.}~\bibnamefont{Bulgac}}, \bibnamefont{and}
	\bibinfo{author}{\bibfnamefont{K.~J.} \bibnamefont{Roche}},
	\bibinfo{journal}{Phys. Rev. A} \textbf{\bibinfo{volume}{88}},
	\bibinfo{pages}{013639} (\bibinfo{year}{2013}).
	
	\bibitem[{\citenamefont{Bluhm and Schafer}(2014)}]{SchaeferPRA14}
	\bibinfo{author}{\bibfnamefont{M.}~\bibnamefont{Bluhm}} \bibnamefont{and}
	\bibinfo{author}{\bibfnamefont{T.}~\bibnamefont{Schafer}},
	\bibinfo{journal}{Phys. Rev. A} \textbf{\bibinfo{volume}{90}},
	\bibinfo{pages}{063615} (\bibinfo{year}{2014}).
	
	\bibitem[{\citenamefont{He and Levin}(2014)}]{YanPRB14}
	\bibinfo{author}{\bibfnamefont{Y.}~\bibnamefont{He}} \bibnamefont{and}
	\bibinfo{author}{\bibfnamefont{K.}~\bibnamefont{Levin}},
	\bibinfo{journal}{Phys. Rev. B} \textbf{\bibinfo{volume}{89}},
	\bibinfo{pages}{035106} (\bibinfo{year}{2014}).
	
	\bibitem[{\citenamefont{Bluhm and Schafer}(2015)}]{SchaferPRA15}
	\bibinfo{author}{\bibfnamefont{M.}~\bibnamefont{Bluhm}} \bibnamefont{and}
	\bibinfo{author}{\bibfnamefont{T.}~\bibnamefont{Schafer}},
	\bibinfo{journal}{Phys. Rev. A} \textbf{\bibinfo{volume}{92}},
	\bibinfo{pages}{043602} (\bibinfo{year}{2015}).
	
	\bibitem[{\citenamefont{Joseph et~al.}(2015)\citenamefont{Joseph, Elliott, and
			Thomas}}]{ThomasPRL15}
	\bibinfo{author}{\bibfnamefont{J.~A.} \bibnamefont{Joseph}},
	\bibinfo{author}{\bibfnamefont{E.}~\bibnamefont{Elliott}}, \bibnamefont{and}
	\bibinfo{author}{\bibfnamefont{J.~E.} \bibnamefont{Thomas}},
	\bibinfo{journal}{Phys. Rev. Lett.} \textbf{\bibinfo{volume}{115}},
	\bibinfo{pages}{020401} (\bibinfo{year}{2015}).
	
	\bibitem[{\citenamefont{Bluhm and Schafer}(2016)}]{SchaeferPRL16}
	\bibinfo{author}{\bibfnamefont{M.}~\bibnamefont{Bluhm}} \bibnamefont{and}
	\bibinfo{author}{\bibfnamefont{T.}~\bibnamefont{Schafer}},
	\bibinfo{journal}{Phys. Rev. Lett.} \textbf{\bibinfo{volume}{116}},
	\bibinfo{pages}{115301} (\bibinfo{year}{2016}).
	
	\bibitem[{\citenamefont{Kovtun et~al.}(2005{\natexlab{b}})\citenamefont{Kovtun,
			Son, and Starinets}}]{TDSonPRL05}
	\bibinfo{author}{\bibfnamefont{P.~K.} \bibnamefont{Kovtun}},
	\bibinfo{author}{\bibfnamefont{D.~T.} \bibnamefont{Son}}, \bibnamefont{and}
	\bibinfo{author}{\bibfnamefont{A.~O.} \bibnamefont{Starinets}},
	\bibinfo{journal}{Phys. Rev. Lett.} \textbf{\bibinfo{volume}{94}},
	\bibinfo{pages}{111601} (\bibinfo{year}{2005}{\natexlab{b}}).
	
	\bibitem[{\citenamefont{Turlapov et~al.}(2008)\citenamefont{Turlapov, Kinast,
			Clancy, Luo, Joseph, and Thomas}}]{Turlapov08}
	\bibinfo{author}{\bibfnamefont{A.}~\bibnamefont{Turlapov}},
	\bibinfo{author}{\bibfnamefont{J.}~\bibnamefont{Kinast}},
	\bibinfo{author}{\bibfnamefont{B.}~\bibnamefont{Clancy}},
	\bibinfo{author}{\bibfnamefont{L.}~\bibnamefont{Luo}},
	\bibinfo{author}{\bibfnamefont{J.}~\bibnamefont{Joseph}}, \bibnamefont{and}
	\bibinfo{author}{\bibfnamefont{J.~E.} \bibnamefont{Thomas}},
	\bibinfo{journal}{J. Low Temp. Phys.} \textbf{\bibinfo{volume}{150}},
	\bibinfo{pages}{567} (\bibinfo{year}{2008}).
	
	\bibitem[{\citenamefont{Cao et~al.}(2011{\natexlab{b}})\citenamefont{Cao,
			Elliott, Wu, and Thomas}}]{Cao11}
	\bibinfo{author}{\bibfnamefont{C.}~\bibnamefont{Cao}},
	\bibinfo{author}{\bibfnamefont{E.}~\bibnamefont{Elliott}},
	\bibinfo{author}{\bibfnamefont{H.}~\bibnamefont{Wu}}, \bibnamefont{and}
	\bibinfo{author}{\bibfnamefont{J.~E.} \bibnamefont{Thomas}},
	\bibinfo{journal}{New J. Phys.} \textbf{\bibinfo{volume}{13}},
	\bibinfo{pages}{075007} (\bibinfo{year}{2011}{\natexlab{b}}).
	
	\bibitem[{\citenamefont{Guo et~al.}(2011{\natexlab{a}})\citenamefont{Guo,
			Wulin, Chien, and Levin}}]{HaoPRL11}
	\bibinfo{author}{\bibfnamefont{H.}~\bibnamefont{Guo}},
	\bibinfo{author}{\bibfnamefont{D.}~\bibnamefont{Wulin}},
	\bibinfo{author}{\bibfnamefont{C.~C.} \bibnamefont{Chien}}, \bibnamefont{and}
	\bibinfo{author}{\bibfnamefont{K.}~\bibnamefont{Levin}},
	\bibinfo{journal}{Phys. Rev. Lett.} \textbf{\bibinfo{volume}{107}},
	\bibinfo{pages}{020403} (\bibinfo{year}{2011}{\natexlab{a}}).
	
	\bibitem[{\citenamefont{Guo et~al.}(2011{\natexlab{b}})\citenamefont{Guo,
			Wulin, Chien, and Levin}}]{HaoNJP11}
	\bibinfo{author}{\bibfnamefont{H.}~\bibnamefont{Guo}},
	\bibinfo{author}{\bibfnamefont{D.}~\bibnamefont{Wulin}},
	\bibinfo{author}{\bibfnamefont{C.~C.} \bibnamefont{Chien}}, \bibnamefont{and}
	\bibinfo{author}{\bibfnamefont{K.}~\bibnamefont{Levin}},
	\bibinfo{journal}{New J. Phys.} \textbf{\bibinfo{volume}{13}},
	\bibinfo{pages}{075011} (\bibinfo{year}{2011}{\natexlab{b}}).
	
	\bibitem[{\citenamefont{Zwierlein
			et~al.}(2006{\natexlab{a}})\citenamefont{Zwierlein, Schirotzek, Schunck, and
			Ketterle}}]{ZSSK06}
	\bibinfo{author}{\bibfnamefont{M.~W.} \bibnamefont{Zwierlein}},
	\bibinfo{author}{\bibfnamefont{A.}~\bibnamefont{Schirotzek}},
	\bibinfo{author}{\bibfnamefont{C.~H.} \bibnamefont{Schunck}},
	\bibnamefont{and} \bibinfo{author}{\bibfnamefont{W.}~\bibnamefont{Ketterle}},
	\bibinfo{journal}{Science} \textbf{\bibinfo{volume}{311}},
	\bibinfo{pages}{492} (\bibinfo{year}{2006}{\natexlab{a}}).
	
	\bibitem[{\citenamefont{Zwierlein
			et~al.}(2006{\natexlab{b}})\citenamefont{Zwierlein, Schunck, Schirotzek, and
			Ketterle}}]{ZSSK206}
	\bibinfo{author}{\bibfnamefont{M.~W.} \bibnamefont{Zwierlein}},
	\bibinfo{author}{\bibfnamefont{C.~H.} \bibnamefont{Schunck}},
	\bibinfo{author}{\bibfnamefont{A.}~\bibnamefont{Schirotzek}},
	\bibnamefont{and} \bibinfo{author}{\bibfnamefont{W.}~\bibnamefont{Ketterle}},
	\bibinfo{journal}{Nature (London)} \textbf{\bibinfo{volume}{442}},
	\bibinfo{pages}{54} (\bibinfo{year}{2006}{\natexlab{b}}).
	
	\bibitem[{\citenamefont{Shin et~al.}(2007)\citenamefont{Shin, Schunck,
			Schirotzek, and Ketterle}}]{ZSSK07}
	\bibinfo{author}{\bibfnamefont{Y.}~\bibnamefont{Shin}},
	\bibinfo{author}{\bibfnamefont{C.~H.} \bibnamefont{Schunck}},
	\bibinfo{author}{\bibfnamefont{A.}~\bibnamefont{Schirotzek}},
	\bibnamefont{and} \bibinfo{author}{\bibfnamefont{W.}~\bibnamefont{Ketterle}},
	\bibinfo{journal}{Nature (London)} \textbf{\bibinfo{volume}{451}},
	\bibinfo{pages}{689} (\bibinfo{year}{2007}).
	
	\bibitem[{\citenamefont{Shin et~al.}(2008)\citenamefont{Shin, Schirotzek,
			Schunck, and Ketterle}}]{MITPRL08}
	\bibinfo{author}{\bibfnamefont{Y.~I.} \bibnamefont{Shin}},
	\bibinfo{author}{\bibfnamefont{A.}~\bibnamefont{Schirotzek}},
	\bibinfo{author}{\bibfnamefont{C.~H.} \bibnamefont{Schunck}},
	\bibnamefont{and} \bibinfo{author}{\bibfnamefont{W.}~\bibnamefont{Ketterle}},
	\bibinfo{journal}{Phys. Rev. Lett.} \textbf{\bibinfo{volume}{101}},
	\bibinfo{pages}{070404} (\bibinfo{year}{2008}).
	
	\bibitem[{\citenamefont{Partridge
			et~al.}(2006{\natexlab{a}})\citenamefont{Partridge, Li, Kamar, Liao, and
			Hulet}}]{Rice06}
	\bibinfo{author}{\bibfnamefont{G.~B.} \bibnamefont{Partridge}},
	\bibinfo{author}{\bibfnamefont{W.}~\bibnamefont{Li}},
	\bibinfo{author}{\bibfnamefont{R.~I.} \bibnamefont{Kamar}},
	\bibinfo{author}{\bibfnamefont{Y.~A.} \bibnamefont{Liao}}, \bibnamefont{and}
	\bibinfo{author}{\bibfnamefont{R.~G.} \bibnamefont{Hulet}},
	\bibinfo{journal}{Science} \textbf{\bibinfo{volume}{311}},
	\bibinfo{pages}{503} (\bibinfo{year}{2006}{\natexlab{a}}).
	
	\bibitem[{\citenamefont{Partridge
			et~al.}(2006{\natexlab{b}})\citenamefont{Partridge, Li, Liao, Hulet, Haque,
			and Stoof}}]{RicePRL06}
	\bibinfo{author}{\bibfnamefont{G.~B.} \bibnamefont{Partridge}},
	\bibinfo{author}{\bibfnamefont{W.}~\bibnamefont{Li}},
	\bibinfo{author}{\bibfnamefont{Y.~A.} \bibnamefont{Liao}},
	\bibinfo{author}{\bibfnamefont{R.~G.} \bibnamefont{Hulet}},
	\bibinfo{author}{\bibfnamefont{M.}~\bibnamefont{Haque}}, \bibnamefont{and}
	\bibinfo{author}{\bibfnamefont{H.~T.~C.} \bibnamefont{Stoof}},
	\bibinfo{journal}{Phys. Rev. Lett.} \textbf{\bibinfo{volume}{97}},
	\bibinfo{pages}{190407} (\bibinfo{year}{2006}{\natexlab{b}}).
	
	\bibitem[{\citenamefont{Chien et~al.}(2006)\citenamefont{Chien, Chen, He, and
			Levin}}]{Chien06}
	\bibinfo{author}{\bibfnamefont{C.-C.} \bibnamefont{Chien}},
	\bibinfo{author}{\bibfnamefont{Q.~J.} \bibnamefont{Chen}},
	\bibinfo{author}{\bibfnamefont{Y.}~\bibnamefont{He}}, \bibnamefont{and}
	\bibinfo{author}{\bibfnamefont{K.}~\bibnamefont{Levin}},
	\bibinfo{journal}{Phys. Rev. Lett.} \textbf{\bibinfo{volume}{97}},
	\bibinfo{pages}{090402} (\bibinfo{year}{2006}).
	
	\bibitem[{\citenamefont{Liao et~al.}(2010)\citenamefont{Liao, Rittner,
			Paprotta, Li, Partridge, Hulet, Baur, and Mueller}}]{Liao10}
	\bibinfo{author}{\bibfnamefont{Y.~A.} \bibnamefont{Liao}},
	\bibinfo{author}{\bibfnamefont{A.~S.~C.} \bibnamefont{Rittner}},
	\bibinfo{author}{\bibfnamefont{T.}~\bibnamefont{Paprotta}},
	\bibinfo{author}{\bibfnamefont{W.}~\bibnamefont{Li}},
	\bibinfo{author}{\bibfnamefont{G.~B.} \bibnamefont{Partridge}},
	\bibinfo{author}{\bibfnamefont{R.~G.} \bibnamefont{Hulet}},
	\bibinfo{author}{\bibfnamefont{S.~K.} \bibnamefont{Baur}}, \bibnamefont{and}
	\bibinfo{author}{\bibfnamefont{E.~J.} \bibnamefont{Mueller}},
	\bibinfo{journal}{Nature} \textbf{\bibinfo{volume}{467}},
	\bibinfo{pages}{567} (\bibinfo{year}{2010}).
	
	\bibitem[{\citenamefont{Chen et~al.}(2005)\citenamefont{Chen, Stajic, Tan, and
			Levin}}]{Ourreview}
	\bibinfo{author}{\bibfnamefont{Q.~J.} \bibnamefont{Chen}},
	\bibinfo{author}{\bibfnamefont{J.}~\bibnamefont{Stajic}},
	\bibinfo{author}{\bibfnamefont{S.~N.} \bibnamefont{Tan}}, \bibnamefont{and}
	\bibinfo{author}{\bibfnamefont{K.}~\bibnamefont{Levin}},
	\bibinfo{journal}{Phys. Rep.} \textbf{\bibinfo{volume}{412}},
	\bibinfo{pages}{1} (\bibinfo{year}{2005}).
	
	\bibitem[{\citenamefont{Nozi\`{e}res and Schmitt-Rink}(1985)}]{NSR}
	\bibinfo{author}{\bibfnamefont{P.}~\bibnamefont{Nozi\`{e}res}}
	\bibnamefont{and}
	\bibinfo{author}{\bibfnamefont{S.}~\bibnamefont{Schmitt-Rink}},
	\bibinfo{journal}{J. Low Temp. Phys.} \textbf{\bibinfo{volume}{59}},
	\bibinfo{pages}{195} (\bibinfo{year}{1985}).
	
	\bibitem[{\citenamefont{Haussmann et~al.}(2007)\citenamefont{Haussmann,
			Rantner, Cerrito, and Zwerger}}]{ZwergerPRA07}
	\bibinfo{author}{\bibfnamefont{R.}~\bibnamefont{Haussmann}},
	\bibinfo{author}{\bibfnamefont{W.}~\bibnamefont{Rantner}},
	\bibinfo{author}{\bibfnamefont{S.}~\bibnamefont{Cerrito}}, \bibnamefont{and}
	\bibinfo{author}{\bibfnamefont{W.}~\bibnamefont{Zwerger}},
	\bibinfo{journal}{Phys. Rev. A} \textbf{\bibinfo{volume}{75}},
	\bibinfo{pages}{023610} (\bibinfo{year}{2007}).
	
	\bibitem[{\citenamefont{Chien et~al.}(2010)\citenamefont{Chien, Guo, He, and
			Levin}}]{OurNSR10}
	\bibinfo{author}{\bibfnamefont{C.~C.} \bibnamefont{Chien}},
	\bibinfo{author}{\bibfnamefont{H.}~\bibnamefont{Guo}},
	\bibinfo{author}{\bibfnamefont{Y.}~\bibnamefont{He}}, \bibnamefont{and}
	\bibinfo{author}{\bibfnamefont{K.}~\bibnamefont{Levin}},
	\bibinfo{journal}{Phys. Rev. A} \textbf{\bibinfo{volume}{81}},
	\bibinfo{pages}{023622} (\bibinfo{year}{2010}).
	
	\bibitem[{\citenamefont{Chien et~al.}(2007)\citenamefont{Chien, Chen, He, and
			Levin}}]{ChienPRL}
	\bibinfo{author}{\bibfnamefont{C.-C.} \bibnamefont{Chien}},
	\bibinfo{author}{\bibfnamefont{Q.~J.} \bibnamefont{Chen}},
	\bibinfo{author}{\bibfnamefont{Y.}~\bibnamefont{He}}, \bibnamefont{and}
	\bibinfo{author}{\bibfnamefont{K.}~\bibnamefont{Levin}},
	\bibinfo{journal}{Phys. Rev. Lett.} \textbf{\bibinfo{volume}{98}},
	\bibinfo{pages}{110404} (\bibinfo{year}{2007}).
	
	\bibitem[{\citenamefont{Guo et~al.}(2013)\citenamefont{Guo, Chien, and
			He}}]{OurJLTP13}
	\bibinfo{author}{\bibfnamefont{H.}~\bibnamefont{Guo}},
	\bibinfo{author}{\bibfnamefont{C.~C.} \bibnamefont{Chien}}, \bibnamefont{and}
	\bibinfo{author}{\bibfnamefont{Y.}~\bibnamefont{He}}, \bibinfo{journal}{J.
		Low Temp. Phys.} \textbf{\bibinfo{volume}{172}}, \bibinfo{pages}{5}
	(\bibinfo{year}{2013}).
	
	\bibitem[{\citenamefont{Guo et~al.}(2017)\citenamefont{Guo, Cai, He, and
			Chien}}]{OurPRA17}
	\bibinfo{author}{\bibfnamefont{H.}~\bibnamefont{Guo}},
	\bibinfo{author}{\bibfnamefont{W.}~\bibnamefont{Cai}},
	\bibinfo{author}{\bibfnamefont{Y.}~\bibnamefont{He}}, \bibnamefont{and}
	\bibinfo{author}{\bibfnamefont{C.~C.} \bibnamefont{Chien}},
	\bibinfo{journal}{Phys. Rev. A} \textbf{\bibinfo{volume}{95}},
	\bibinfo{pages}{033638} (\bibinfo{year}{2017}).
	
	\bibitem[{\citenamefont{Guo et~al.}(2012)\citenamefont{Guo, Chien, and
			He}}]{OurPRD12}
	\bibinfo{author}{\bibfnamefont{H.}~\bibnamefont{Guo}},
	\bibinfo{author}{\bibfnamefont{C.~C.} \bibnamefont{Chien}}, \bibnamefont{and}
	\bibinfo{author}{\bibfnamefont{Y.}~\bibnamefont{He}}, \bibinfo{journal}{Phys.
		Rev. D} \textbf{\bibinfo{volume}{85}}, \bibinfo{pages}{074025}
	(\bibinfo{year}{2012}).
	
	\bibitem[{\citenamefont{Pao et~al.}(2006)\citenamefont{Pao, Wu, and Yip}}]{Pao}
	\bibinfo{author}{\bibfnamefont{C.~H.} \bibnamefont{Pao}},
	\bibinfo{author}{\bibfnamefont{S.~T.} \bibnamefont{Wu}}, \bibnamefont{and}
	\bibinfo{author}{\bibfnamefont{S.~K.} \bibnamefont{Yip}},
	\bibinfo{journal}{Phys. Rev. B} \textbf{\bibinfo{volume}{73}},
	\bibinfo{pages}{132506} (\bibinfo{year}{2006}).
	
	\bibitem[{\citenamefont{Pieri and Strinati}(2006)}]{StrinatiPRL06}
	\bibinfo{author}{\bibfnamefont{P.}~\bibnamefont{Pieri}} \bibnamefont{and}
	\bibinfo{author}{\bibfnamefont{G.~C.} \bibnamefont{Strinati}},
	\bibinfo{journal}{Phys. Rev. Lett.} \textbf{\bibinfo{volume}{96}},
	\bibinfo{pages}{150404} (\bibinfo{year}{2006}).
	
	\bibitem[{\citenamefont{Liu and Hu}(2006)}]{HuiHuEPL06}
	\bibinfo{author}{\bibfnamefont{X.~J.} \bibnamefont{Liu}} \bibnamefont{and}
	\bibinfo{author}{\bibfnamefont{H.}~\bibnamefont{Hu}},
	\bibinfo{journal}{Europhys. Lett.} \textbf{\bibinfo{volume}{75}},
	\bibinfo{pages}{364} (\bibinfo{year}{2006}).
	
	\bibitem[{\citenamefont{Leggett}(1980)}]{Leggett}
	\bibinfo{author}{\bibfnamefont{A.~J.} \bibnamefont{Leggett}}, in
	\emph{\bibinfo{booktitle}{Modern Trends in the Theory of Condensed Matter}}
	(\bibinfo{publisher}{Springer-Verlag}, \bibinfo{address}{Berlin},
	\bibinfo{year}{1980}), pp. \bibinfo{pages}{13--27}.
	
	\bibitem[{\citenamefont{Chien}(2009)}]{Chienthesis}
	\bibinfo{author}{\bibfnamefont{C.~C.} \bibnamefont{Chien}},
	\emph{\bibinfo{title}{Ph.D. Thesis}} (\bibinfo{publisher}{University of
		Chicago}, \bibinfo{year}{2009}).
	
	\bibitem[{\citenamefont{Levin et~al.}(2010)\citenamefont{Levin, Chen, Chien,
			and He}}]{OurAnnPhys}
	\bibinfo{author}{\bibfnamefont{K.}~\bibnamefont{Levin}},
	\bibinfo{author}{\bibfnamefont{Q.~J.} \bibnamefont{Chen}},
	\bibinfo{author}{\bibfnamefont{C.~C.} \bibnamefont{Chien}}, \bibnamefont{and}
	\bibinfo{author}{\bibfnamefont{Y.}~\bibnamefont{He}}, \bibinfo{journal}{Ann.
		Phys.} \textbf{\bibinfo{volume}{325}}, \bibinfo{pages}{233}
	(\bibinfo{year}{2010}).
	
	\bibitem[{Cal()}]{Caldas03_04}
	\bibinfo{note}{P. F. Bedaque, H. Caldas, and G. Rupak, Phys. Rev. Lett.
		\textbf{91}, 247002 (2003); H. Caldas, Phys. Rev. A \textbf{69}, 063602
		(2004).}
	
	\bibitem[{\citenamefont{Sarma}(1963)}]{Sarma}
	\bibinfo{author}{\bibfnamefont{G.}~\bibnamefont{Sarma}}, \bibinfo{journal}{J.
		Phys. Chem. Solids,} \textbf{\bibinfo{volume}{24}}, \bibinfo{pages}{1029}
	(\bibinfo{year}{1963}).
	
	\bibitem[{\citenamefont{Kadanoff and Martin}(1961)}]{Kadanoff61}
	\bibinfo{author}{\bibfnamefont{L.~P.} \bibnamefont{Kadanoff}} \bibnamefont{and}
	\bibinfo{author}{\bibfnamefont{P.~C.} \bibnamefont{Martin}},
	\bibinfo{journal}{Phys. Rev.} \textbf{\bibinfo{volume}{124}},
	\bibinfo{pages}{670} (\bibinfo{year}{1961}).
	
	\bibitem[{\citenamefont{Nambu}(1960)}]{Nambu60}
	\bibinfo{author}{\bibfnamefont{Y.}~\bibnamefont{Nambu}},
	\bibinfo{journal}{Phys. Rev.} \textbf{\bibinfo{volume}{117}},
	\bibinfo{pages}{648} (\bibinfo{year}{1960}).
	
	\bibitem[{\citenamefont{Chen et~al.}(2007)\citenamefont{Chen, He, Chien, and
			Levin}}]{LOFFlong}
	\bibinfo{author}{\bibfnamefont{Q.~J.} \bibnamefont{Chen}},
	\bibinfo{author}{\bibfnamefont{Y.}~\bibnamefont{He}},
	\bibinfo{author}{\bibfnamefont{C.~C.} \bibnamefont{Chien}}, \bibnamefont{and}
	\bibinfo{author}{\bibfnamefont{K.}~\bibnamefont{Levin}},
	\bibinfo{journal}{Phys. Rev. B} \textbf{\bibinfo{volume}{75}},
	\bibinfo{pages}{014521} (\bibinfo{year}{2007}).
	
	\bibitem[{\citenamefont{Conti and Vignale}(1999)}]{Conti}
	\bibinfo{author}{\bibfnamefont{S.}~\bibnamefont{Conti}} \bibnamefont{and}
	\bibinfo{author}{\bibfnamefont{G.}~\bibnamefont{Vignale}},
	\bibinfo{journal}{Phys. Rev. B} \textbf{\bibinfo{volume}{60}},
	\bibinfo{pages}{7966} (\bibinfo{year}{1999}).
	
	\bibitem[{\citenamefont{Dorfle et~al.}(1980)\citenamefont{Dorfle, Brand, and
			Graham}}]{Dorfle80}
	\bibinfo{author}{\bibfnamefont{M.}~\bibnamefont{Dorfle}},
	\bibinfo{author}{\bibfnamefont{H.}~\bibnamefont{Brand}}, \bibnamefont{and}
	\bibinfo{author}{\bibfnamefont{R.}~\bibnamefont{Graham}},
	\bibinfo{journal}{J. Phys. C} \textbf{\bibinfo{volume}{13}},
	\bibinfo{pages}{3337} (\bibinfo{year}{1980}).
	
	\bibitem[{\citenamefont{Massignan et~al.}(2005)\citenamefont{Massignan, Bruun,
			and Smith}}]{BruunPRA05}
	\bibinfo{author}{\bibfnamefont{P.}~\bibnamefont{Massignan}},
	\bibinfo{author}{\bibfnamefont{G.~M.} \bibnamefont{Bruun}}, \bibnamefont{and}
	\bibinfo{author}{\bibfnamefont{H.}~\bibnamefont{Smith}},
	\bibinfo{journal}{Phys. Rev. A} \textbf{\bibinfo{volume}{71}},
	\bibinfo{pages}{033607} (\bibinfo{year}{2005}).
	
	\bibitem[{\citenamefont{Fetter and Walecka}(2003)}]{Walecka}
	\bibinfo{author}{\bibfnamefont{A.~L.} \bibnamefont{Fetter}} \bibnamefont{and}
	\bibinfo{author}{\bibfnamefont{J.~D.} \bibnamefont{Walecka}},
	\emph{\bibinfo{title}{Quantum Theory of Many-Particle Systems}}
	(\bibinfo{publisher}{Dover Publications}, \bibinfo{year}{2003}).
	
	\bibitem[{\citenamefont{Horikoshi et~al.}(2016)\citenamefont{Horikoshi, Koashi,
			Tajima, Ohashi, and Kuwata-Gonokami}}]{Horikoshi17}
	\bibinfo{author}{\bibfnamefont{M.}~\bibnamefont{Horikoshi}},
	\bibinfo{author}{\bibfnamefont{M.}~\bibnamefont{Koashi}},
	\bibinfo{author}{\bibfnamefont{H.}~\bibnamefont{Tajima}},
	\bibinfo{author}{\bibfnamefont{Y.}~\bibnamefont{Ohashi}}, \bibnamefont{and}
	\bibinfo{author}{\bibfnamefont{M.}~\bibnamefont{Kuwata-Gonokami}},
	\emph{\bibinfo{title}{Ground-state thermodynamic quantities of homogeneous
			spin-1/2 fermions from the bcs region to the unitarity limit}}
	(\bibinfo{year}{2016}), \bibinfo{note}{arXiv: 1612.04026}.
	
	\bibitem[{\citenamefont{Mukherjee et~al.}(2017)\citenamefont{Mukherjee, Yan,
			Patel, Hadzibabic, Yefsah, Struck, and Zwierlein}}]{Mukherjee17}
	\bibinfo{author}{\bibfnamefont{B.}~\bibnamefont{Mukherjee}},
	\bibinfo{author}{\bibfnamefont{Z.}~\bibnamefont{Yan}},
	\bibinfo{author}{\bibfnamefont{P.~B.} \bibnamefont{Patel}},
	\bibinfo{author}{\bibfnamefont{Z.}~\bibnamefont{Hadzibabic}},
	\bibinfo{author}{\bibfnamefont{T.}~\bibnamefont{Yefsah}},
	\bibinfo{author}{\bibfnamefont{J.}~\bibnamefont{Struck}}, \bibnamefont{and}
	\bibinfo{author}{\bibfnamefont{M.~W.} \bibnamefont{Zwierlein}},
	\bibinfo{journal}{Phys. Rev. Lett.} \textbf{\bibinfo{volume}{118}},
	\bibinfo{pages}{123401} (\bibinfo{year}{2017}).
	
	\bibitem[{\citenamefont{Guo et~al.}(2014)\citenamefont{Guo, Li, He, and
			Chien}}]{OurJPB14}
	\bibinfo{author}{\bibfnamefont{H.}~\bibnamefont{Guo}},
	\bibinfo{author}{\bibfnamefont{Y.}~\bibnamefont{Li}},
	\bibinfo{author}{\bibfnamefont{Y.}~\bibnamefont{He}}, \bibnamefont{and}
	\bibinfo{author}{\bibfnamefont{C.}~\bibnamefont{Chien}}, \bibinfo{journal}{J.
		Phys. B: At. Mol. Opt. Phys.} \textbf{\bibinfo{volume}{47}},
	\bibinfo{pages}{085302} (\bibinfo{year}{2014}).
	
	\bibitem[{\citenamefont{Schrieffer}(1964)}]{Schrieffer_book}
	\bibinfo{author}{\bibfnamefont{J.~R.} \bibnamefont{Schrieffer}},
	\emph{\bibinfo{title}{Theory of superconductivity}}
	(\bibinfo{publisher}{Benjamin}, \bibinfo{address}{New York},
	\bibinfo{year}{1964}).
	
	\bibitem[{\citenamefont{Kadanoff and Martin}(1963)}]{KM2}
	\bibinfo{author}{\bibfnamefont{L.~P.} \bibnamefont{Kadanoff}} \bibnamefont{and}
	\bibinfo{author}{\bibfnamefont{P.~C.} \bibnamefont{Martin}},
	\bibinfo{journal}{Annals of Physics} \textbf{\bibinfo{volume}{24}},
	\bibinfo{pages}{419} (\bibinfo{year}{1963}).
	
\end{thebibliography}

\end{document}